\documentclass[prd,eqsecnum,showpacs,twocolumn]{revtex4}
\usepackage[tbtags]{amsmath}
\usepackage{amsfonts}
\usepackage{amssymb}
\usepackage[final]{graphicx}
\usepackage{hyperref}
\usepackage{bm}
\allowdisplaybreaks[1]

\newcommand{\paperI}{paper~I}
\newcommand{\paperII}{paper~II}

\renewcommand{\vec}[1]{\bm{#1}}
\newcommand{\conj}[1]{\bar{#1}}
\newcommand{\Dirac}{\delta}
\newcommand{\Heaviside}{\theta}
\newcommand{\approaches}{\rightarrow}
\renewcommand{\d}{\mathrm{d}}
\newcommand{\odiff}[2]{\frac{\d#1}{\d#2}}
\newcommand{\diff}[2]{\frac{\partial#1}{\partial#2}}
\newcommand{\diffz}[2]{\frac{\partial^2#1}{\partial{#2}^2}}
\newcommand{\diffzm}[3]{\frac{\partial^2#1}{\partial#2\partial#3}}
\newcommand{\programname}[1]{\texttt{\textsc{#1}}}
\newcommand{\routinename}[1]{\texttt{#1}}
\newcommand{\abs}[1]{\left\lvert{#1}\right\rvert}

\newcommand{\jump}[1]{\left[#1\right]}
\newcommand{\ccp}[1]{\gamma^{#1}}
\newcommand{\ccm}[1]{\epsilon^{#1}}
\newcommand{\orderof}[1]{O(#1)}
\newcommand{\real}{\operatorname{Re}}
\newcommand{\imag}{\operatorname{Im}}
\newcommand{\define}{\equiv}

\newcommand{\sign}{\operatorname{sign}}
\newcommand{\updntensor}[3]{#1^{#2}_{\phantom{#2}#3}}
\newcommand{\pp}[2]{\updntensor{g}{#1}{#2}}
\newcommand{\Chr}[2]{\updntensor{\Gamma}{#1}{#2}}
\newcommand{\Riemann}[2]{\updntensor{R}{#1}{#2}}
\newcommand{\A}[2]{\updntensor{A}{#1}{#2}}
\newcommand{\B}[2]{\updntensor{B}{#1}{#2}}
\newcommand{\radv}{r_{\text{adv}}}
\newcommand{\frakf}{\mathfrak{f}}
\newcommand{\eg}{\mbox{e.g.\@}}

\begin{document}
%%fakesection
% author information
\title{Time domain calculation of the electromagnetic self-force on eccentric
geodesics in Schwarzschild spacetime}
\author{Roland \surname{Haas}}
\affiliation{Theoretical Astrophysics,
California Institute of Technology, Pasadena, CA 91125
}
\affiliation{Center for Relativistic Astrophysics,
Georgia Institute of Technology, Atlanta, GA 30332
}
\date{\today}
% 02.60.Cb : Numerical simulation; solution of equations

%abstract
\begin{abstract}
    I calculate the self-force acting on a  particle with electric charge $q$
    moving on a generic geodesic around a Schwarzschild black hole. 
    Using methods similar to those developed for the scalar field case
    discussed
    in~\cite{haas:07}, I investigate the relative sizes of the conservative
    (half-advanced plus half-retarded) and dissipative (half-advanced minus
    half-retarded) pieces of the self-force. I also display the
    regularization parameters used in the mode-sum regularization scheme.
\end{abstract}

\maketitle

% ---------------------- main document --------------------------------------
\section{Introduction\label{sec:introduction}}
This is the second paper of a series of papers studying the self-force on a
point particle in generic geodesic orbit around a Schwarzschild black hole. 
I extend the previous calculation of the scalar self-force~\cite{haas:06}
to electromagnetism,
studying in particular the effects of the conservative part of the self-force. 

A test particle in orbit around a black hole will follow a geodesic. Going
beyond the test mass limit, this is no longer true and the particle's path will
deviate from a geodesic of the background spacetime. 
As seen from the background spacetime, the particle is said
to experiences a self-force due to its interaction with its own field. 
In order to accurately model the motion of the body, including its
inspiral toward the black hole, I seek to evaluate the self-force and
calculate its effect on the motion. Several methods to achieve this have been
proposed in the literature~\cite{barack:00, vega:07, Barack:2007we}. 
I elect to use the mode-sum regularization scheme introduced by 
Barack and Ori~\cite{barack:00}, which been proven to be highly accurate.

In this paper, rather than dealing with the gravitational problem, I 
focus on the technically simpler problem of a point particle
endowed with an electric charge $q$ orbiting a Schwarzschild black hole of
mass $M$. In this context I use a numerical simulation to check the
analytically calculated regularization parameters used in the mode-sum
regularization scheme, which I calculate in a manner analogous
to~\cite{haas:06}. This calculation also makes it possible to investigate the
behaviour of the conservative (half-advanced plus half-retarded) part of the
self-force in the strong-field limit, extending previous work by Pound and
Poisson~\cite{pound:07}. Different from the scalar case calculation, where the
conservative self-force is suppressed, the conservative electromagnetic 
self-force appears at the same post Newtonian order as the gravitational
conservative self-force.
Agreement, even if only qualitative, between the results for the 
electromagnetic problem, where our physical intuition allows us to understand the
mechanisms at work, and those for a point mass recently explored
by~\cite{barack:2011ed,Warburton:2011fk} can thus help provide a clearer
understanding of the
mechanisms at work in the gravitational case as well.

Throughout the paper I use geometrized units in which $G = c = 1$ and the
sign conventions of~\cite{MTW:73}.

\subsection{The problem\label{sec:the-problem}}
Since my approach is essentially identical to that described
in~\cite{haas:06} and \cite{haas:07}
(\paperI{} and \paperII{} from now on), 
I will only briefly introduce the required notation.

The first order self-force is calculated on a geodesic of Schwarzschild
spacetime, whose metric is written in Schwarzschild coordinates as
\begin{equation}
    \d s^2 = -f\d t^2 + f^{-1}\d r^2 + r^2\d\Omega\text{,}
\end{equation}
where $f = \left( 1-\frac{2M}{r} \right)$, $\d\Omega = \left( \d \theta^2 +
\sin^2\theta \d\phi^2 \right)$ is the metric on a two-sphere, and $t$, $r$, 
$\theta$ and $\phi$ are the usual Schwarzschild coordinates.
I numerically solve the Maxwell equations
\begin{gather}
    g^{\beta\gamma} \nabla_\gamma  F_{\alpha\beta}(x) = 4\pi j_\alpha(x)
    \text{,}\label{eqn:covariant-mw-eqn}
    \\
    \nabla_{[\gamma}  F_{\alpha\beta]}(x) = 0
    \text{,}\label{eqn:covariant-mw-constraint}
    \\
    j_\alpha(x) = q \int_\gamma u_\alpha(\tau) \Dirac_4\bm(x,z(\tau)\bm) \d\tau\text{,}
\end{gather}
where $\nabla_\alpha$ is the covariant derivative compatible with the metric
$g_{\alpha\beta}$, $F_{\alpha\beta}$ is the Faraday field tensor sourced by a charge
$q$ which moves along a world line $\gamma: \tau \mapsto z(\tau)$ parametrized by 
proper time
$\tau$. The current density $j_\alpha(x)$ appearing on the right-hand side
is written in terms of a scalarized four-dimensional Dirac $\Dirac$-function
$\Dirac_4(x,x') \define \Dirac(x_0-x'_0) \Dirac(x_1-x'_1) \Dirac(x_2-x'_2)
\Dirac(x_3-x'_3)/\sqrt{-\det(g_{\alpha\beta})}$.
After having obtained the  Faraday tensor 
I regularize it using the mode-sum
regularization scheme introduced by Barack and Ori~\cite{barack:00}
\begin{multline}
    F^R_{(\mu)(\nu)} = F^{\text{ret}}_{(\mu)(\nu)} 
      - q \sum_\ell \left[
          A_{(\mu)(\nu)} \Bigl(\ell + \frac 12\Bigr) 
	+ B_{(\mu)(\nu)} 
	\right.\\\mbox{}\left.
	+ \frac{C_{(\mu)(\nu)}}{\ell + \frac 12}
        + \frac{D_{(\mu)(\nu)}}{(\ell - \frac 12) (\ell + \frac 32)} 
	+ \cdots \right] \text{,}\label{eqn:mode-by-mode-regularization}
\end{multline}
where indices in parenthesis $(\mu)$ signify components with respect to an
orthonormal tetrad $e^\alpha_{\ (\mu)}$ and 
the coefficients $A_{(\mu)(\nu)}$, $B_{(\mu)(\nu)}$, $C_{(\mu)(\nu)}$,
and $D_{(\mu)(\nu)}$ are independent of $\ell$; they are listed in
Appendix~\ref{sec:regularization-parameters}. Finally I compute the
regularized self-force 
\begin{equation}
    F^{\text{self}}_\alpha \define q F^R_{\alpha\beta} u^\beta
    \label{eqn:self-force-from-faraday-tensor}
\end{equation}
from the regularized Faraday tensor and the four velocity of the particle.

\subsection{Organization of this paper}
In Sec.~\ref{sec:numerical-method} I introduce the ideas behind the
discretization scheme used in the numerical simulation.
Sec.~\ref{sec:initial-and-boundary-condition} describes the choices I make
in order to handle the problems of specifying initial data and proper boundary
conditions.  In
Sec.~\ref{sec:numerical-tests} I describe the tests I performed in order
to validate my implementation of the numerical method.
Sec.~\ref{sec:sample-results} contains 
sample results for a small number of representative simulations. Finally in 
Sec.~\ref{sec:conservative-self-force} I calculate the conservative
self-force for the same set of simulations. The appendices contain technical
details and an alternative calculation using the vector potential instead of
the Faraday tensor. 

\section{Numerical method\label{sec:numerical-method}}
In this section I describe the algorithm used to integrate the Maxwell
equations numerically. I
use the second-order algorithm introduced by Lousto and Price~\cite{lousto:97}
suitably extended to handle a coupled system of equations.

\subsection{Wave equations for the Faraday
tensor\label{sec:wave-eqn-for-faraday}}
I introduce a vector potential $A_\alpha$ in terms of which the Faraday
tensor is given by
\begin{align}
    F_{\alpha\beta} &= A_{\beta,\alpha} -
    A_{\alpha,\beta}\text{,}\label{eqn:faraday-from-vector-potential}
\end{align}
where a comma denotes an ordinary derivative.
I use vector spherical harmonics 
$Z^{\ell m}_A = \partial_A Y^{\ell m}$
and $X^{\ell m}_A = {\epsilon_A}^B \partial_B Y^{\ell m}$, where
$\epsilon_{AB}$ is the Levi-Civita tensor associated with the metric
$\Omega_{AB}$ on the two-sphere ($\epsilon_{\theta\phi} = \sin\theta$), as
introduced in~\cite{regge:57, martel:05}. I decompose the vector potential
and the current density into
\begin{subequations}
\begin{align}
    A_a(t,r,\theta,\phi) &= A^{\ell m}_a(t, r) Y_{\ell  m}(\theta, \phi)
    \text{,}
    \\
    j_a(t,r,\theta,\phi) &= j^{\ell m}_a(t, r) Y_{\ell m}(\theta, \phi)
      & \text{for $a = t,r$,}
    \\
    A_A(t,r,\theta,\phi) &= v_{\ell m}(t, r) Z_A^{\ell m}(\theta, \phi) 
    \nonumber\\&\quad
    + \tilde v_{\ell m}(t, r) X_A^{\ell m}(\theta, \phi)
    \text{,}
    \\
    j_A(t,r,\theta,\phi) &= j^{\text{even}}_{\ell m}(t, r) Z_A^{\ell m}(\theta, \phi) 
    \nonumber\\&\quad
    + j^{\text{odd}}_{\ell m}(t, r) X_A^{\ell m}(\theta, \phi)
      & \text{for $A = \theta,\phi$,}
\end{align}
\end{subequations}
where a summation over $\ell$ and $m$ is implied. Substituting these into
Eq.~\eqref{eqn:covariant-mw-eqn} I arrive at two sets of coupled
equations for the even ($A^{\ell m}_a$, $v_{\ell m}$) and odd ($\tilde v_{\ell
m}$) modes
\begin{subequations}
\begin{gather}
    -f \diffz{A_t^{\ell m}}{r} + f \diffzm{A_r^{\ell m}}{t}{r}
    -\frac{2f}{r} \diff{A_t^{\ell m}}{r} + \frac{2f}{r} \diff{A_r^{\ell m}}{t}
    \nonumber \\ \mbox{}-\frac{\ell(\ell+1)}{r^2} \diff{v_{\ell m}}{t} 
    +\frac{\ell(\ell+1)}{r^2} A_t^{\ell m} = 4\pi j_t^{\ell m}
    \text{,}\label{eqn:mw-t}
    \\
    f^{-1} \diffz{A_r^{\ell m}}{t} - f^{-1} \diffzm{A_t^{\ell m}}{t}{r}
    -\frac{\ell(\ell+1)}{r^2} \diff{v_{\ell m}}{r} \nonumber \\ \mbox{}
    +\frac{\ell(\ell+1)}{r^2} A_r^{\ell m} = 4\pi j_r^{\ell m}
    \text{,}\label{eqn:mw-r}
    \\
    f^{-1} \diffz{v_{\ell m}}{t} - f \diffz{v_{\ell m}}{r}
    -\frac{2M}{r^2} \diff{v_{\ell m}}{r}
    +f \diff{A_r^{\ell m}}{r}  \nonumber \\ \mbox{} - f^{-1} \diff{A_t^{\ell m}}{t}
    +\frac{2M}{r^2} A_r^{\ell m} = 4\pi j^{\text{even}}_{\ell m}
    \text{,}\label{eqn:mw-v}
    \\
    f^{-1} \diffz{\tilde v_{\ell m}}{t} - f \diffz{\tilde v_{\ell m}}{r}
    -\frac{2M}{r^2} \diff{\tilde v_{\ell m}}{r} \nonumber \\ \mbox{}
    +\frac{\ell(\ell+1)}{r^2} \tilde v_{\ell m} = 4\pi j^{\text{odd}}_{\ell m}
    \text{,}\label{eqn:mw-vtilde}
\end{gather}
\end{subequations}
where
\begin{subequations}
\label{eqn:red-sources}
\begin{gather}
    j^{\ell m}_t = -\frac{q f}{r_0^2} 
      \conj Y^{\ell m}(\frac\pi2, \varphi_0) \Dirac(r-r_0)
      \text{,}
    \\
    j^{\ell m}_r = \frac{q \dot r_0}{E r_0^2} 
      \conj Y^{\ell m}(\frac\pi2, \varphi_0) \Dirac(r-r_0) 
      \text{,}
    \\
    j^{\text{even}}_{\ell m} = -\frac{i m q f J}{\ell(\ell+1) E r_0^2} 
      \conj Y^{\ell m}(\frac\pi2, \varphi_0) \Dirac(r-r_0)
      \text{,}
    \\
    j^{\text{odd}}_{\ell m} = -\frac{q f J}{\ell(\ell+1) E r_0^2} 
      \partial_\theta \conj Y^{\ell m}(\frac\pi2, \varphi_0) \Dirac(r-r_0) 
      \text{.}
\end{gather}
\end{subequations}
In the equation above an overbar denotes complex
conjugation, an overdot denotes differentiation with respect to $\tau$,
$E = -u_t$ is the particle's conserved energy per unit mass,
$J = u_\phi$ its conserved angular momentum per unit mass,
and $u^\alpha = \odiff{z^\alpha}{\tau}$ is its four velocity.
Quantities bearing a subscript ``$0$'' are evaluated at the
particle's position; they are functions of $\tau$ that are obtained by solving
the geodesic equation 
\begin{equation}
    u^\beta \nabla_\beta u^\alpha = 0
\end{equation}
in the background spacetime.
Without loss of generality, I have confined the motion of the
particle to the equatorial plane $\theta = \frac\pi2$. 

The three even mode equations Eq.~\eqref{eqn:mw-t} -- Eq.~\eqref{eqn:mw-v}
are not yet amenable to a numerical treatment, as they are highly coupled. In
order to obtain a more convenient set of equation I define the auxiliary
fields
\begin{gather}
    \psi^{\ell m} \define -r^2 \left( \diff{A_t^{\ell m}}{r} - \diff{A_r^{\ell m}}{t} \right)
    \text{,}\label{eqn:auxiliary-fields-def}
    \\
    \chi^{\ell m} \define f\, \left( A_r^{\ell m} - \diff{v^{\ell m}}{r} \right)
    \text{,}
    \\
    \xi^{\ell m} \define A_t^{\ell m} - \diff{v^{\ell
    m}}{t}\label{eqn:auxiliary-fields-def-xi}
    \text{,}
\end{gather}
which, up to scaling factors, are just the even multipole moments of the 
$tr$, $r\phi$ and $t\phi$ components of the Faraday tensor
\begin{align}
    F_{tr} &= \sum_{\ell,m} \frac{\psi^{\ell m}}{r^2} \, Y^{\ell
    m}\text{,}\label{eqn:faraday-from-auxiliary-fields-begin}\\
    F_{tA} &= \sum_{\ell,m} (-\xi^{\ell m} \, Z^{\ell m}_A + \tilde v^{\ell m}_{,t} \, X^{\ell m}_A)\text{,}\\
    F_{rA} &= \sum_{\ell,m} (\frac{\chi^{\ell m}}{f} \, Z^{\ell m}_A + \tilde v^{\ell m}_{,r} \, X^{\ell m}_A)\text{,}\\
    F_{\theta\varphi} &= 
    \sum_{\ell,m} \tilde v_{\ell m} \, (X^{\ell m}_{\phi,\theta} - X^{\ell m}_{\theta,\phi})
    \nonumber\\
    &= -\sum_{\ell,m} \ell(\ell+1) \tilde v_{\ell m} \, \sin(\theta) Y^{\ell m}
    \text{.}\label{eqn:faraday-from-auxiliary-fields-end}
\end{align}

I note that the three fields $\psi^{\ell m}$, $\chi^{\ell m}$
and $\xi^{\ell m}$ are not independent of each other, in fact knowledge of $\psi^{\ell m}$ is
sufficient to reconstruct $\chi^{\ell m}$ and $\xi^{\ell m}$. Eq.~\eqref{eqn:mw-t} can be
rearranged to yield
\begin{gather}
    \xi^{\ell m} = -\frac{f}{\ell(\ell+1)} \diff{\psi^{\ell m}}{r} - \frac{4\pi}{\ell(\ell+1)} j_t^{\ell m}
    \text{,}\label{eqn:xi-from-psi}\\
    \intertext{and similarly from Eq.~\eqref{eqn:mw-r}}
    \chi^{\ell m} = -\frac{1}{\ell(\ell+1)} \diff{\psi^{\ell m}}{t} - \frac{4\pi
    f}{\ell(\ell+1)} j_r^{\ell m}
    \text{,}\label{eqn:chi-from-psi}
\end{gather}
showing that knowledge of $\psi^{\ell m}$ is sufficient to reconstruct the even
multipole components of the Faraday tensor. In this work however I choose to
solve for $\chi^{\ell m}$ and $\xi^{\ell m}$ directly, rather than to
numerically differentiate $\psi^{\ell m}$ to obtain them. The gain in speed
from reducing the number of equations
does not seem to offset the additional time required to calculate $\psi^{\ell m}$
accurately enough to obtain good approximations for its derivatives at the
location of the particle. In this approach Eqs.~\eqref{eqn:xi-from-psi}
and~\eqref{eqn:chi-from-psi} are treated as constraints that the
dynamical variables have to satisfy. 

Dropping the superscripts $\ell$, $m$ for notational convenience and 
following~\cite{cunningham:79} I form linear combinations of derivatives of Eqs.\eqref{eqn:mw-t}
-- \eqref{eqn:mw-v}. I use 
$[\partial_r (r^2\, \text{\eqref{eqn:mw-r}}) - 
\partial_t (r^2 \, \text{\eqref{eqn:mw-t}}) ]$ 
for $\psi$ and find
\begin{subequations}
\begin{gather}
    \diffz{\psi}{r^*} - \diffz{\psi}{t} - V \psi = S_\psi \text{,}\\
    S_\psi = 
    4\pi f \left[ \diff{(r^2 j_t^{\ell m})}{r} - \diff{(r^2 j_r^{\ell m})}{t} \right]
    \text{,}\label{eqn:red-wave-psi}
\end{gather}
where $V = \ell(\ell+1) \frac{r-2M}{r^3}$ and $r^* = r +
2M\ln(\frac{r}{2M}-1)$ is the Regge-Wheeler tortoise coordinate.
Similarly I use 
$[f \, \text{\eqref{eqn:mw-r}} - \partial_r (f \, \text{\eqref{eqn:mw-v}}) ]$ 
for $\chi$ and 
$[\text{\eqref{eqn:mw-t}} - \partial_t \text{\eqref{eqn:mw-v}} ]$ 
for $\xi$. I find 
\begin{align}
    \diffz{\chi}{r^*} - \diffz{\chi}{t} - V \chi
    &=  S_\chi \text{,}\\
    S_\chi = 4\pi f \Biggl[ \diff{(f j^{\text{even}}_{\ell m})}{r} &- f j_r^{\ell m} \Biggr]
    \text{,}\label{eqn:red-wave-chi}
    \\
    \diffz{\xi}{r^*} - \diffz{\xi}{t} - V \xi
    - V_\xi \psi &= S_\xi \text{,}\\
    S_\xi = 4\pi f 
    \Biggl[ \diff{(f j^{\text{even}}_{\ell m})}{t} &- f j_t^{\ell m} \Biggr]
    \text{,}\label{eqn:red-wave-xi}
\end{align}
\end{subequations}
where $V_\xi = \frac{2(r-3M)(r-2M)}{r^5}$.
While still partially coupled 
Eqs.~\eqref{eqn:red-wave-psi} -- \eqref{eqn:red-wave-xi} are much easier to
deal with than the original set Eqs.~\eqref{eqn:mw-t} -- \eqref{eqn:mw-v}. 
The coupling is in the form of a staggering, which allows me
to first solve for $\psi$ and use this result in the calculation of $\xi$. On
the other hand, the source terms appearing on the right-hand side contain
derivatives of Dirac's $\Dirac$-function resulting in fields that are 
discontinuous at the
location of the particle. Lousto's scheme is designed to cope with
precisely this situation. 

I derive explicit expressions for the source terms $S_\alpha$ on the
right hand sides
\begin{subequations}
\begin{gather}
    S_\alpha = G_\alpha(t) f_0 \Dirac(r-r_0) + F_\alpha(t) f \Dirac'(r-r_0)
    \text{,}\label{eqn:source-term-schematic}
    \\
    G_\psi(t) = -\frac{4\pi q}{E^2} f_0 \, \left(
      \ddot r_0 - \frac{i m \dot r_0 J}{r_0^2} \right) 
      \bar Y_{\ell m}(\frac{\pi}{2}, \varphi_0)
    \text{,}
    \\
    F_\psi(t) = 4\pi q f_0 \left( \frac{\dot r_0^2}{E^2} - 1\right) 
      \bar Y_{\ell m}(\frac{\pi}{2}, \varphi_0)
    \text{,}
    \\
    G_\chi(t) = -\frac{4\pi q \dot r_0}{E r_0^2} f_0 
    \bar Y_{\ell m}(\frac{\pi}{2}, \varphi_0)
    \text{,}
    \\
    F_\chi(t) = -\frac{4\pi q J i m}{E \ell(\ell+1) r_0^2} f_0^2 
    \bar Y_{\ell m}(\frac{\pi}{2}, \varphi_0)
    \text{,}
    \\
    \begin{split}
    G_\xi(t) &= -4\pi q \biggl\{
    \frac{J i m}{E^2 \ell(\ell+1) r_0^2} \biggl[ \biggl(
    \frac{2M}{r_0^2}-\frac{2f_0}{r_0} \biggr) \dot r_0 
    \nonumber\\&\qquad
    - \frac{imJ}{r_0^2} \biggr]
    -\frac{1}{r_0^2} \biggr\} f_0 \bar Y_{\ell m}(\frac{\pi}{2}, \varphi_0)
    \text{,}
    \end{split}
    \\
    F_\xi(t) = \frac{4\pi q J i m \dot r_0}{E^2 \ell(\ell+1) r_0^2} f_0^2 \bar
    Y_{\ell m}(\frac{\pi}{2}, \varphi_0)
    \text{.}
\end{gather}
\end{subequations}
My functions $G_\alpha$ and $F_\alpha$ correspond to $G/f_0$ and $F/f$ 
in~\cite{lousto:97},
respectively, they  are independent
of $r$ (but do contain terms in $r_0(t)$).
I prefer this form of the
source terms over the form given in~\cite{lousto:97} since it simplifies
the integral over the source term Eq.~(3.6) of~\cite{lousto:97}
\begin{multline}
    \iint \d A S = 2 \int_{t_1}^{t_2} 
    \biggl[ \frac{G\bm(r_0(t),t\bm)}{1-2M/r_0(t)} 
     \\
    - \diff{}{r}
     \biggl(\frac{F(r,t)}{1-2M/r}\biggr)\bigg|_{r = r_0(t)}
    \biggr] \d t
    \\
    \pm 2 \frac{F\bm(r_0(t_1),t_1\bm)}{[1-2M/r_0(t_1)]^2} [1\mp\dot r^*_0(t_1)]^{-1}
    \\
    \pm 2 \frac{F\bm(r_0(t_2),t_2\bm)}{[1-2M/r_0(t_2)]^2} [1\pm\dot r^*_0(t_2)]^{-1}
    \text{.}\label{eqn:lousto-source-integral}
\end{multline}
Since $G^{\text{Lousto}} = f_0 G_\alpha(t)$ and $G^{\text{Lousto}} = f
F_\alpha(t)$,
the first term in square brackets inside the integral simplifies, while the 
second term vanishes completely. $F_\alpha$ only appears in the boundary
terms.

\subsection{Constraint equations\label{sec:constraint-equations}}
The full set of Maxwell equations consists of the inhomogeneous equations
Eq.~\eqref{eqn:covariant-mw-eqn} as well as the homogeneous constraints
Eq.~\eqref{eqn:covariant-mw-constraint}
which have to be satisfied by a solution to Eq.~\eqref{eqn:covariant-mw-eqn}.
In the usual approach introducing a vector potential $A_\alpha$ implies that the constraints are
identically satisfied since they reduce to the Bianchi identities for the
second derivatives of $A_\alpha$. When solving for the components of the Faraday
tensor directly there is no a priory guarantee that a solution to
Eq.~\eqref{eqn:red-wave-psi} -- \eqref{eqn:red-wave-xi},
and~\eqref{eqn:mw-vtilde} satisfies Eq.~\eqref{eqn:covariant-mw-constraint}.
It turns out, however, that a decomposition into spherical harmonics 
is
sufficient to show that all but one of the constraints are identically
satisfied. The one that is not identically true is the $tr\varphi$ (or
$tr\theta$) equation, which in terms of $\psi$, $\chi$ and $\xi$ reads
\begin{equation}
    \frac{\psi}{r^2} - \frac{\chi_{,t}}{f} + \xi_{,r} = 0
    \text{.}\label{eqn:trphi-constraint}
\end{equation}
If the fields satisfy the sourced Maxwell equations Eqs.~\eqref{eqn:mw-t},
\eqref{eqn:mw-r},
then Eq.~\eqref{eqn:trphi-constraint} is just the evolution equation for
$\psi$. Thus Eq.~\eqref{eqn:trphi-constraint} is valid whenever $\psi$
satisfies the consistency relations Eq.~\eqref{eqn:xi-from-psi} 
and~\eqref{eqn:chi-from-psi}.

Analytically then, the situation is clear. Given a set of compatible 
initial conditions
for $\psi$, $\chi$ and $\xi$ which initially satisfy the constraint equations, 
a solution to the system of Eq.~\eqref{eqn:red-wave-psi} -- 
\eqref{eqn:red-wave-xi}, \eqref{eqn:mw-vtilde} satisfies the full set of
Maxwell equations at
all later times, too. 

\begin{figure}
    \includegraphics{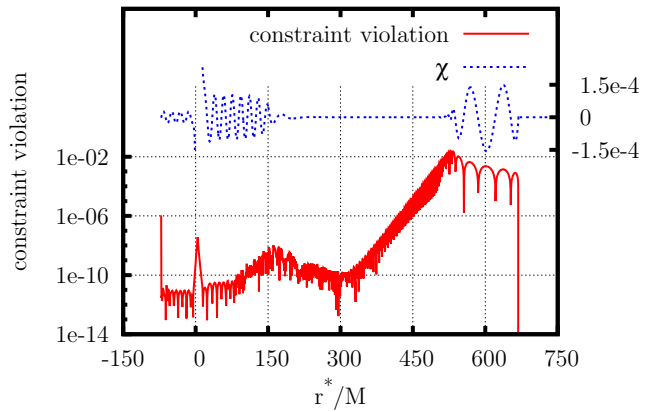}
    \caption{Violations of the constraint 
    $Z_\chi = \chi + \frac{1}{\ell(\ell+1)} \diff{\psi}{t} = 0$
    in the vacuum region away from the location of particle. I plot the
    $\chi$ and $\log_{10}\abs{Z_\chi}$ as
    obtained on a spatial slice at time $t=600\,M$. 
    For this slightly eccentric orbit ($p = 7.0$, $e = 0.3$) using a stepsize
    $h = 1/512M$ the
    errors in the $\ell = 2$, $m = 2$ mode are at least three orders of
    magnitude smaller than the field values. The exponentially growing signal
    between $300M \lesssim r^* \lesssim 500$ is a remnant of the initial data
    pulse travelling outward. 
    \label{fig:constraint-violations}}
\end{figure}

\begin{figure}
    \includegraphics{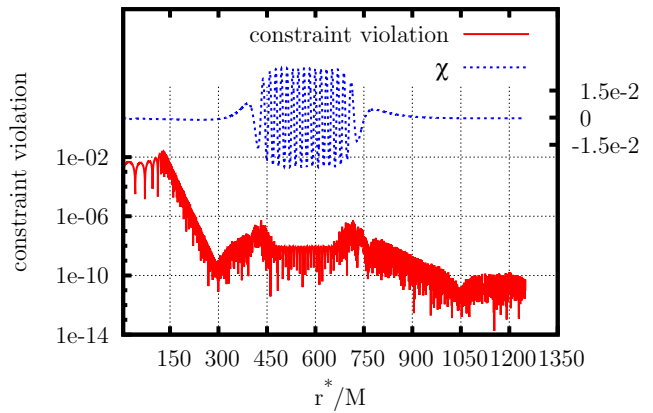}
    \caption{Violations of the constraint 
    $Z_\chi = \chi + \frac{1}{\ell(\ell+1)} \diff{\psi}{t} = 0$
    at the location of the particle as a function of time. 
    I display $\chi$ and $\log_{10}\abs{Z_\chi}$ for
    the $\ell = 5$, $m = 3$ mode of a particle on an eccentric orbit with 
    $p = 7.8001$, $e = 0.9$ with stepsize $h = 1/256\,M$.
    During the time $400\,M \lesssim t \lesssim
    800\,M$ the particle is in the whirl phase. The exponentially decaying
    signal before $t \approx 250M$ is the initial data pulse.
    \label{fig:zoom-whirl-constraint-violations}}
\end{figure}

Numerically I monitor but do not enforce Eq.~\eqref{eqn:xi-from-psi} 
and~\eqref{eqn:chi-from-psi}. I generally find that violations of the
constraints are at least three orders of magnitude smaller than the field
quantities themselves. Figures~\ref{fig:constraint-violations}
and~\ref{fig:zoom-whirl-constraint-violations} compare
$\chi$ obtained from its evolution equations to that obtained
from Eq.~\eqref{eqn:chi-from-psi}.

\subsection{Monopole mode\label{sec:monopole-mode}}
For the electromagnetic field, the monopole mode $\ell = 0$ is
non-radiative. The vector harmonics $Z^{\ell m}_A$ and $X^{\ell m}_A$ cannot
be defined in this case and the only surviving multipole mode is  $\psi$.
For the monopole case Eq.~\eqref{eqn:red-wave-psi} reduces to a wave equation
in flat space
\begin{equation}
    \diffz{\psi}{r^*} - \diffz{\psi}{t} = 4\pi
    f \left[ \diff{(r^2 j_t^{0,0})}{r} - \diff{(r^2 j_r^{0,0})}{t} \right]
    \text{,}\label{eqn:red-wave-monopole}
\end{equation}
which is simple enough so that I can solve it analytically. A straightforward
calculation shows that 
\begin{equation}
    \psi(t, r^*) = -\sqrt{4\pi} q \Heaviside(r^* - r^*_0(t))
    \label{eqn:monopole-solution}
\end{equation}
satisfies Eq.~\eqref{eqn:red-wave-monopole} and corresponds to no 
outgoing radiation $(\partial_t-\partial_{r^*}) \psi = 0$
at the event horizon and no ingoing radiation $(\partial_t+\partial_{r^*}) \psi = 0$
at spatial
infinity. 

\subsection{Discretization---even sector}
\begin{figure}
    \includegraphics{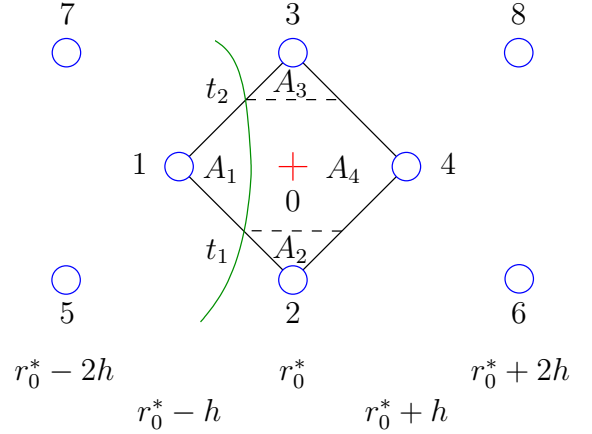}
    \caption{Points used to calculate the integral over the potential terms.
    Grid points are indicated by blue circles.}
    \label{fig:simpson-points}
\end{figure}

Lousto's method is directly applicable to terms of the form
$-\diffz{\psi}{t}+\diffz{\psi}{r^*}$, $V(r) \psi$ (ie.\ the wave operator
and potential terms) on the left-hand side of the equation 
and the source terms $S_\alpha(t)$
on the right hand side.
Here $\psi$ is used as a placeholder for any one of $\psi$, $\chi$ or $\xi$;
$V(r)$ is an expression depending only on $r$.
I discretize these as
\begin{gather}
    \iint_{\text{\rlap{cell}}} \, \d u \, \d v \, \left(
    -\diffz{\psi}{t}+\diffz{\psi}{r^*} \right) = 
    -4\left[\psi_3 + \psi_2 - \psi_1 - \psi_4\right]
    \text{,}\label{eqn:differential-operator}
    \\
    \iint_{\text{\rlap{cell}}} \, \d u \, \d v \,  V(r)\psi = 
    \begin{cases}
	h^2 V_0 \, \sum_i \psi_i + \orderof{h^4} & \text{vacuum cells} \\
	V_0 \, \sum_i A_i \psi_i +	\orderof{h^3} & \text{sourced cells,}
    \end{cases}
    \label{eqn:pot-integral}
\end{gather}
and
\begin{align}
    \iint_{\text{\rlap{cell}}} \, \d u \, \d v \, S_\alpha(t) &= 
    2\int_{t_1}^{t_2} G_\alpha\bm(t, r_0(t)\bm) \, \d t  
     \nonumber \\ & \qquad 
    \pm \frac{2 F_\alpha\bm(t_1, r_0(t_1)\bm)}{1-2M/r(t_1)} [1\mp\dot r_0(t_1)/E]^{-1}
     \nonumber \\ & \qquad 
    \pm \frac{2 F_\alpha\bm(t_2, r_0(t_2)\bm)}{1-2M/r(t_2)} [1\pm\dot r_0(t_2)/E]^{-1}
    \text{,}\label{eqn:source-integral}
\end{align}
where $u=t-r^*$, $v=t+r^*$ are null coordinates, $\psi_1$,\ldots,$\psi_4$
refer to values of the field at the points labelled $1$,\ldots,$4$ in
Fig.~\ref{fig:simpson-points}, $h=\Delta_t = \Delta_{r^*}/2$ is the step size,
$V_0$ is the value of the potential at the centre of the cell,
$A_1$,\ldots,$A_4$ are the areas indicated in Fig.~\ref{fig:simpson-points} 
and $t_1$ and
$t_2$ are the times at which the particle enters and leaves the cell, respectively.

Spelled out explicitly the evolution equations for vacuum cells are
\begin{subequations}
\begin{align}
    \psi_3 &= -\psi_2 +  (1-\frac{h^2}{2} V_0) (\psi_1 + \psi_4) \text{,}\\
    \chi_3 &= -\chi_2 +  (1-\frac{h^2}{2} V_0) (\chi_1 + \chi_4) \text{,}\\
    \xi_3 &= -\xi_2 +  (1-\frac{h^2}{2} V_0) (\xi_1 + \xi_4) 
    \nonumber \\&\quad
    -\frac{h^2}{4} V_{\xi,0} (\psi_1+\psi_2+\psi_3+\psi_4)
    \text{,}
\end{align}
\end{subequations}
and for sourced cells
\begin{subequations}
\begin{align}
    \psi_3 &= -[1 + \frac{V_0}{4}(A_2-A_3)]\psi_2 
      + [1-\frac{V_0}{4}(A_4+A_3)]\psi_4 
      \nonumber\\&\quad
      + [1-\frac{V_0}{4}(A_1+A_3)]\psi_1
      \nonumber\\&\quad
      - \frac14 (1-\frac{V_0}{4} A_3) \iint \d u \,\d v\, S_\psi(t)
      \text{,}\\
    \chi_3 &= -[1 + \frac{V_0}{4}(A_2-A_3)]\chi_2 
      + [1-\frac{V_0}{4}(A_4+A_3)]\chi_4 
      \nonumber\\&\quad
      + [1-\frac{V_0}{4}(A_1+A_3)]\chi_1
      \nonumber\\&\quad
      - \frac14 (1-\frac{V_0}{4} A_3) \iint \d u \,\d v\, S_\chi(t)
      \text{,}\\
    \xi_3 &= -[1 + \frac{V_0}{4}(A_2-A_3)]\xi_2 
      + [1-\frac{V_0}{4}(A_4+A_3)]\xi_4 
      \nonumber\\&\quad
      + [1-\frac{V_0}{4}(A_1+A_3)]\xi_1
      \nonumber\\&\quad
      - \frac14 V_{\xi,0} (A_1\psi_1+A_2\psi_2+A_3\psi_3+A_4\psi_4)
      \nonumber\\&\quad
      - \frac14 (1-\frac{V_0}{4} A_3) \iint \d u \,\d v\, S_\xi(t)
    \text{.}
\end{align}
\end{subequations}

\subsection{Discretization--odd sector}
When written in terms of $r^*$, Eq.~\eqref{eqn:mw-vtilde}, which governs the odd
modes $\tilde v^{\ell m}$, is
\begin{gather}
    \diffz{\tilde v_{\ell m}}{r^*} - \diffz{\tilde v_{\ell m}}{t}
    -\frac{\ell(\ell+1)(r-2M)}{r^3} \tilde v_{\ell m} = -4\pi f j^{\text{odd}}_{\ell m}
    \text{,\label{eqn:red-wave-vtilde}}
    \\
    j^{\text{odd}}_{\ell m} = -\frac{q J}{\ell(\ell+1) E r_0^2} 
      \partial_\theta \conj Y^{\ell m}(\frac\pi2, \varphi_0) \Dirac(r^*-r^*_0) 
    \text{.}
\end{gather}
Eq.~\eqref{eqn:red-wave-vtilde} is of the form of the scalar
wave equation discussed in \paperII{}. I re-use the fourth order numerical
code described there with $V = \frac{\ell(\ell+1)(r-2M)}{r^3}$, $S = 4\pi
\frac{q f J}{\ell(\ell+1) E r_0^2} \partial_\theta 
\conj Y^{\ell m}(\frac\pi2, \varphi_0)$. This yields accurate results for
$\tilde v$ and its derivatives. 

\section{Initial values and boundary conditions\label{sec:initial-and-boundary-condition}}
I follow the approach detailed in \paperII{} for the
scalar self-force and do not specify physical initial data or an
outer boundary condition.
I arbitrarily choose the fields to vanish on the characteristic
slices $u = u_0 = t_0 - r^*_0$ and $v = v_0 = t_0 + r^*_0$
\begin{equation}
    \psi(u = u_0) = \psi(v=v_0) = 0 \text{,}
\end{equation}
thereby adding a certain amount of spurious waves to the solution which show
up as an initial burst.

I implement ingoing wave boundary conditions near the event horizon,
sufficiently close that \emph{numerically} $r \approx 2M$, so that the potential
terms in Eqs.~\eqref{eqn:red-wave-psi} -- \eqref{eqn:red-wave-xi}
vanish. This happens at
$r^* \approx -73\,M$ and I implement the ingoing waves condition 
$\partial_u \psi = 0$ there. Near the outer boundary this is not
possible, since the potential decays slowly. Instead I choose to evolve
the full domain of dependence of the initial data surface, hiding the
effects of the boundary.

\section{Particle motion\label{sec:particle-motion}}
I use the same approach as described in~\paperII{} to evolve the
particle's motion, i.e.\ I introduce the semi-latus rectum $p$, the 
eccentricity $e$ and a fictitious angle $\chi$, not to be confused with the
Faraday tensor component $\chi$ defined in
Eq.~\eqref{eqn:auxiliary-fields-def-xi}, such that 
\begin{equation}
    r(\tau) = \frac{pM}{1+e \cos \chi(\tau)}\text{.}
\end{equation}
The evolution is then governed by
\begin{gather}
    \begin{split}
    \odiff{\chi}{t} =
    \frac{(p-2-2e\cos\chi)(1+e\cos\chi)^2}{(Mp^2)}
    \\\mbox{}\times
    \sqrt{ \frac{p-6-2e\cos\chi}{(p-2-2e)(p-2+2e)} }\text{,}
    \end{split}
    \\
    \odiff{\varphi}{t} =
    \frac{(p-2-2e\cos\chi)(1+e\cos\chi)^2}{p^{3/2}M\sqrt{(p-2-2e)(p-2+2e)}}
    \text{.}\label{eqn:particle-ode}
\end{gather}
I use the embedded Runge-Kutta-Fehlberg (4, 5) algorithm provided by the
\programname{GNU Scientific Library} routine
\routinename{gsl\_odeiv\_step\_rkf45} and an adaptive step-size control to
evolve the position of the particle forward in time. 

\section{Extraction of field data at the particle\label{sec:field-extraction}}
I use a straightforward one-sided extrapolation of field values to the right
of the particle's position to extract values for $\psi$ and $\partial_{r^*} \psi$.
Specifically I fit a fourth order polynomial 
\begin{equation}
    p(x)  = 
    \sum_{n=0}^4 \frac{c_i}{n!} x^n \text{,}
\end{equation}
where $x = r^* - r^*_0$ to the five points to the right of the particle's
current position and extract $\psi$ and $\partial_{r^*} \psi$ as $c_0$ and
$c_1$, respectively. 
In order to calculate
$\diff{\psi(t_0, r^*_0)}{t}$ I follow~\cite{sago:07} and calculate
$\odiff{\psi\bm(t, r^*(t)\bm)}{t}$ on the world line of the particle. Since
this can be calculated using either the field values on the world line 
\begin{multline}
    \odiff{\psi\bm(t, r^*(t)\bm)}{t} = 
    \\\mbox{}
    \frac{\psi\bm(t+h, r^*(t+h)\bm) - \psi\bm(t-h, r^*(t-h)\bm)}{2h} 
    + \orderof{h^2}
    \text{,}
\end{multline}
or as 
\begin{equation}
    \odiff{\psi\bm(t, r^*(t)\bm)}{t} = 
	\diff{\psi}{t} + \diff{\psi}{r^*} \odiff{r^*_0}{t}
    \text{,}
\end{equation}
where both  $\diff{\psi}{r^*}$ and $\odiff{r^*_0}{t} = \dot r_0/E$ are known, this allows me
to find
\begin{equation}
    \diff{\psi}{t} = \odiff{\psi\bm(t, r^*(t)\bm)}{t} 
	- \diff{\psi}{r^*} \odiff{r^*_0}{t}\text{.}
\end{equation}

I repeat this procedure to the left of the particle. 
As a check for the extraction procedure, I compare the difference between the
right hand and left hand values $\jump{\psi} = \psi_{\text{right}} -
\psi_{\text{left}}$ with the analytically calculated jump conditions of
appendix~\ref{sec:jump-conditions}. 
Similarly I check whether the numerical solutions obtained for $\chi$ and
$\xi$ directly are consistent with Eqs.~\eqref{eqn:chi-from-psi}
and~\eqref{eqn:xi-from-psi}, which give them in terms of derivatives of
$\psi$.

\section{Regularization of the mode sum}
The regularization procedure operates on scalar spherical harmonic modes of
the multipole coefficients $F^{\ell m}_{(\mu)(\nu)}$ of the Faraday tensor. As a
first step I use the auxiliary fields $\psi$, $\chi$ and $\xi$ 
to reconstruct
\begin{subequations}
\begin{gather}
    A^{\ell'm'}_{r,t}-A^{\ell' m'}_{t,r} = \frac{\psi}{r^2}\text{,}
    \\
    \partial_t v^{\ell'm'}-A^{\ell' m'}_t = -\xi \text{,}
    \\
    \intertext{and}
    \partial_r v^{\ell'm'}-A^{\ell' m'}_r = -\frac{\chi}{f}\text{,}
\end{gather}
\end{subequations}
the combinations of the vector potential modes needed to obtain the even
sector of a tensor spherical harmonic decomposition of the Faraday tensor. 
The auxiliary field $\tilde v$ and its derivatives provide the odd sector of
the decomposition. 

%Next I introduce a pseudo-Cartesian tetrad
%\begin{subequations}
%\label{eqn:tetrad-def-reals}
%\begin{gather} 
%    e^\alpha_{\ (0)} = \biggl[ \frac{1}{\sqrt{f}}, 0, 0, 0
%    \biggr]\text{,}\label{eqn:tetrad-def-0} \\ 
%    e^\alpha_{\ (1)} = \biggl[ 0, \sqrt{f}\sin\theta\cos\phi, 
%    \frac{1}{r} \cos\theta\cos\phi, -\frac{\sin\phi}{r\sin\theta}
%    \biggr]\text{,}\\ 
%    e^\alpha_{\ (2)} = \biggl[ 0, \sqrt{f}\sin\theta\sin\phi, 
%    \frac{1}{r} \cos\theta\sin\phi, \frac{\cos\phi}{r\sin\theta} 
%    \biggr]\text{,}\\ 
%    e^\alpha_{\ (3)} = \biggl[ 0, \sqrt{f}\cos\theta, 
%    -\frac{1}{r} \sin\theta, 0 \biggr]
%\end{gather} 
%\end{subequations}
%together with the complex combinations 
%$e^\alpha_{\ (\pm)} \define e^\alpha_{\ (1)} \pm i e^\alpha_{\ (2)}$,
%\begin{equation} 
%e^\alpha_{\ (\pm)} = \biggl[ 0, \sqrt{f}\sin\theta e^{\pm i \phi},   
%\frac{1}{r} \cos\theta e^{\pm i\phi}, 
%\frac{\pm i e^{\pm i\phi}}{r\sin\theta}
%\biggr]\text{.}\label{eqn:tetrad-def-pm}  
%\end{equation}
%I use these to define tetrad components 
Using the comples pseudo-Cartesian tetrad $e^\alpha_{\ (0)}$, $e^\alpha_{\
(\pm)}$ and $e^\alpha_{\ (3)}$ introduced in \paperI, I define tetrad
components
\begin{equation}
    F^{\text{ret}}_{(\mu)(\nu)} \define F^{\text{ret}}_{\alpha\beta}
    e^\alpha_{\ (\mu)} e^\beta_{\ (\nu)}
\end{equation}
of the Faraday tensor. 

I construct the spherical harmonic modes
of $F^{\text{ret}}_{(\mu)(\nu)}$ using the coupling coefficients displayed in 
Eq.~\eqref{eqn:translation-table}. 
\begin{multline}
    F^{\ell m, \text{ret}}_{(\mu)(\nu)}
    = \sum_{\ell', m'} \left[
      C^{a b}_{(\mu)(\nu)}(\ell' m' | \ell m) \left( A^{\ell'm'}_{b,a}-A^{\ell' m'}_{a,b} \right) 
      \right.\nonumber\\\left.
      + D^a_{(\mu)(\nu)}(\ell' m' | \ell m) \left( \partial_a v^{\ell'm'}-A^{\ell' m'}_a \right)  
      \right.\nonumber\\\left.
      + E^a_{(\mu)(\nu)}(\ell' m' | \ell m) \partial_a \tilde v^{\ell' m'} 
      \right.\nonumber\\\left.
      + E_{(\mu)(\nu)}(\ell' m' | \ell m) \tilde v^{\ell' m'}
      \right]
\end{multline}
I
calculate the multipole coefficients of $F^{\ell,\text{ret}}_{(\mu)(\nu)}$ as
\begin{equation}
    F^{\ell,\text{ret}}_{(\mu)(\nu)} = \sum_m F^{\ell m,\text{ret}}_{(\mu)(\nu)}(t,r_0) 
    Y_{\ell m}(\frac\pi2, \varphi_0)\text{,}
\end{equation}
and regularize them as in Eq.~\eqref{eqn:mode-by-mode-regularization}. 
\begin{multline}
    F^R_{(\mu)(\nu)} = \sum_\ell \biggl\{ F^{\ell,\text{ret}}_{(\mu)(\nu)} - q
    \Bigl[ A_{(\mu)(\nu)} \Bigl(\ell+\frac12\Bigr) + B_{(\mu)(\nu)} +
    \\
    \frac{C_{(\mu)(\nu)}}{\ell+\frac12} +
    \frac{D_{(\mu)(\nu)}}{(\ell-\frac12)(\ell+\frac32)} \Bigr] \biggr\}
\end{multline}
I calculate the regularized self-force using 
$F^R_{(\mu)} = q F^R_{(\mu)(\nu)} u^{(\nu)}$.
Finally I reconstruct the vector components of the self-force 
by from the tetrad components
\begin{subequations}
\begin{gather}
    F^R_t      = \sqrt{f_0} F^R_{(0)}\text{,}
    \\
    F^R_r      = \frac{1}{\sqrt{f_0}} 
    \real \left( F^R_{(+)} e^{-i\varphi_0} \right)\text{,}
    \\
    F^R_\phi   = r_0 \imag \left( F^R_{(+)} e^{-i\varphi_0} \right)\text{.}
\end{gather}
\end{subequations}

\section{Numerical tests\label{sec:numerical-tests}}
In this section I present the tests I performed to validate
my numerical evolution code. I performed the same set of tests as
described in \paperII{}. First, in order to check the second-order
convergence rate of the code, I performed regression runs with increasing
resolution. As a second test, I computed the regularized self-force for several
different combinations of orbital elements $p$ and $e$ and checked that the
multipole coefficients decay with $\ell$ as expected. This provided a very
sensitive check on the overall implementation of the numerical scheme as well
as the analytical calculations that lead to the regularization parameters.

\subsection{Convergence tests\label{sec:convergence-tests-particle}}
\begin{figure}
    \includegraphics{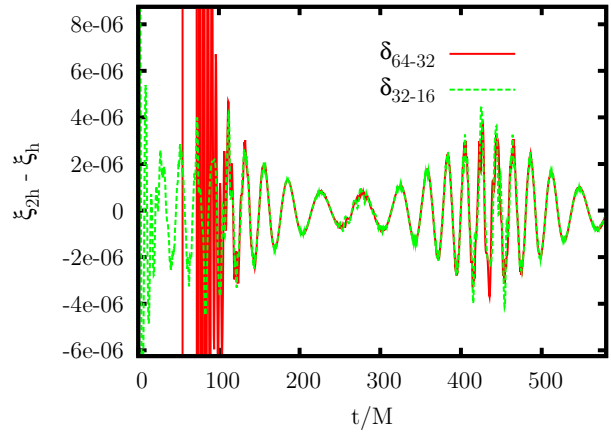}
    \caption{Convergence test of the numerical algorithm in the sourced case. I
    show differences between simulations using different step sizes of 
    16, 32 and 64 cells per $M$.
    Displayed are the rescaled differences $\delta_{32-16} = \xi(h = 1/32 M) -
    \xi(h = 1/16 M)$ etc.\ of the field values at the position of
    the particle for a simulation with $\ell = 6$, $m = 4$ and $p = 7$, $e = 0.3$.  
    I see that the convergence is approximately second-order. The curves are
    rescaled in such a way as to provide an estimate for the error of the
    highest resolution run compared to the real ($h \equiv 0$) solution.}
    \label{fig:sourced-regression}
\end{figure}

Convergence tests are a 
straightforward way to test the implementation of a numerical scheme. I
performed regression runs for my second-order convergent code using 
a non-zero charge $q$ and an eccentric orbit. I extract 
the field at the position of the particle, and thus also test the
implementation of the extraction algorithm described in
section~\ref{sec:field-extraction}. 
I choose the $\ell = 6$, $m = 4$ mode of the field generated by
a particle on a mildly eccentric
geodesic orbit with $p = 7$, $e = 0.3$. As shown in
Fig.~\ref{fig:sourced-regression} the convergence is approximately of second
order.
In the region $150\,M \lesssim t \lesssim 400\,M$ the two curves lie on top of
each other, as expected for a second-order convergent algorithm. In the region
from $400\,M$ to $450\,M$ there is some difference between the two lines,
caused by cell crossing effects similar to those discussed in~\paperII.

\subsection{Discontinuity across the world
line\label{sec:discontinuity-across-world-line}}
The singular source term on the right hand side of
Eqs.~\eqref{eqn:red-wave-psi} -- \eqref{eqn:red-wave-xi} implies that the
fields $\psi$, $\chi$ and $\xi$ are discontinuous across the world line. Since
the jump conditions can be calculated analytically as done in
appendix~\ref{sec:jump-conditions}, I can check whether the numerical results
faithfully reproduce the expected behaviour. Using the methods described in
section~\ref{sec:field-extraction} I obtain one-sided extrapolation for the
field values and their spatial derivatives. For the highest resolution run
used in the regression analysis in
section~\ref{sec:convergence-tests-particle} I find that the numerical
results for $\xi$ agree with the analytical calculation of the jump condition up to
terms of the order of $10^{-8}$; 
% plot "elmag_faraday/sourced-regression_32.out" using 1:($20+$42-$6) with lines
two orders of magnitude smaller than the estimated numerical error of
$10^{-6}$.
For $\partial_{r^*} \xi$ the situation is reversed, with the numerical error in
the jump condition being about an order of magnitude larger than the numerical
error in the field derivative itself. 
% gawk -f even/regression.awk 1 14 2 $files >sourced-regression.dat 
% plot "elmag_faraday/sourced-regression_32.out" using 1:($28+$44-$14) with lines
The accuracy of the numerical derivatives is therefore limited by the accuracy
of the extraction scheme, resulting in about three significant figures for the
set of parameters displayed in Fig.~\ref{fig:sourced-regression}. However the
regularization calculation
is constructed in such a way that no derivatives of the fields need be
obtained in order to calculate the self-force. I therefore feel that I can
accept the reduced accuracy provided by the simple extraction scheme. 

\subsection{High-$\ell$ behaviour of the multipole coefficients\label{sec:high-ell-behavior}}
\begin{figure}
    \includegraphics{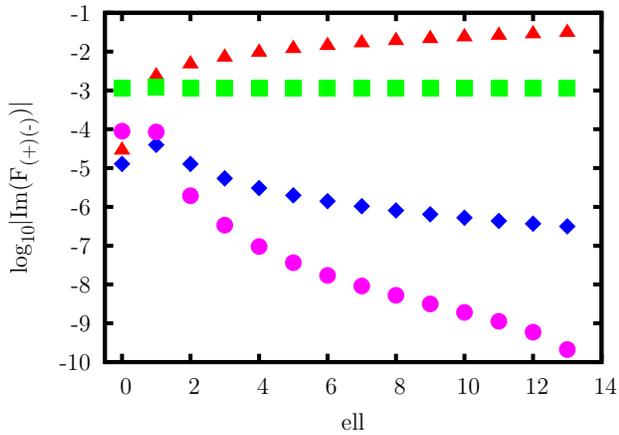}
    \caption{Multipole coefficients of the dimensionless Faraday tensor
    component $\frac{M^2}{q} \imag F^R_{(+)(-)}$ for a
    particle on an eccentric orbit ($p = 7.2$, $e = 0.5$).
    The
    coefficients are extracted at $t = 500\,M$ along the trajectory shown in 
    Fig.~\ref{fig:eccentric-orbit-trajectory}.
    The plots show several stages of the
    regularization procedure, with a closer
    description of the curves to be found in the
    text. A uniform stepsize of $h = 1/512\,M$ was used.
    \label{fig:multipole-coeffs-eccentric-orbit-plusminus}}
\end{figure}

Inspection of Eq.~\eqref{eqn:mode-by-mode-regularization} reveals that a plot of
$F^\ell_{(\mu)(\nu)}$ as a function of $\ell$ (for a fixed value of $t$)
should display a linear growth in $\ell$ for large $\ell$. Removing the
$A_{(\mu)(\nu)}$ term should produce a constant curve, removing the
$B_{(\mu)(\nu)}$ term (given that $C_{(\mu)(\nu)} = 0$) should produce a 
curve that decays as
$\ell^{-2}$, and finally, removing the $D_{(\mu)(\nu)}$ term should 
produce a curve
that decays as $\ell^{-4}$. It is a powerful test of the overall
implementation to
check whether the numerical data behaves as expected. 
Fig.~\ref{fig:multipole-coeffs-eccentric-orbit-plusminus}
plots the remainders as obtained from my numerical simulation,
demonstrating the expected behaviour.
It displays, on a logarithmic scale, the absolute value of 
$\imag F^{\ell,R}_{(+)(-)}$, the imaginary part of the $F^R_{(+)(-)}$ tetrad component of
the Faraday tensor.
The orbit is eccentric ($p =7.2$, $e = 0.5$), and all components of the self-force 
require regularization. The
first curve (in triangles) shows the unregularized multipole coefficients that
increase linearly in $\ell$, as confirmed by fitting 
a straight line to the data. 
% gnuplot:
% Final set of parameters            Asymptotic Standard Error
% =======================            ==========================
% 
% a               = 1.01306          +/- 0.007277     (0.7184%)
% b               = -1.29723         +/- 0.005955     (0.4591%)
The second curve (in squares) shows partially regularized coefficients,
obtained after
the removal of $(\ell+1/2)A_{(\mu)(\nu)}$; this  clearly approaches a constant for large
values of $\ell$. The curve made up of diamonds shows the behaviour after
removal of $B_{(\mu)(\nu)}$; because 
$C_{(\mu)(\nu)} = 0$, it decays as $\ell^{-2}$, a behaviour that is confirmed by a fit to
the $\ell \ge 5$ part of the curve.
% gnuplot:
% Final set of parameters            Asymptotic Standard Error
% =======================            ==========================
% 
% a               = -2.12213         +/- 0.02353      (1.109%)
% b               = -2.03871         +/- 0.02222      (1.09%)
Finally, after removal of $D_{(\mu)(\nu)}/[(\ell - \frac12)\,(\ell + \frac32)]$ the
terms of the sum decrease in magnitude approximately as
$\ell^{-4}$ when fitting to the data points $\ell \ge 7$.
This result depends slightly on the range of points used for the fit.
I expect this to be due to the fact that I stop
at $\ell = 15$, which seems to be not large enough to show the
asymptotic behaviour. Extending the range to very high values of $\ell$
proved to be very difficult, since the numerical code is only second order
convergent, so that the numerical errors become dominant by the time the
asymptotic behaviour begins to show. 
% gnuplot
% Final set of parameters            Asymptotic Standard Error
% =======================            ==========================
% a               = -4.01033         +/- 0.08891      (2.217%)
% b               = -1.10104         +/- 0.08847      (8.035%)

Each one of the last two curves would result in a
converging sum, but the convergence is faster after subtracting the
$D_{(\mu)(\nu)}$ terms. I thereby gain about one order of magnitude in the
accuracy of the estimated sum. 

Figure~\ref{fig:multipole-coeffs-eccentric-orbit-plusminus} provides a sensitive
test of
the implementation
of both the numerical and analytical parts of the calculation. Small
mistakes in either one will cause the difference in
Eq.~\eqref{eqn:mode-by-mode-regularization} to have a vastly different behaviour.

\subsection{Accuracy of the numerical method\label{sec:numerical-accuracy}}
In this work I are less demanding with the numerical accuracy then I were
in~\paperII, where I describe a very high accuracy numerical code.
Implementing suach a code is very tedious even for the scalar case, and much more
so for the electromagnetic case treated here. Therefore I implement a simpler
method that allows me to access the physics of the problem without being
hindered by technical problems due to a complicated numerical method. 

An estimate for the truncation error arising from cutting short the summation
in
Eq.~\eqref{eqn:mode-by-mode-regularization} at some $\ell_{\text{max}}$ can be calculated
by considering the behaviour of the remaining terms for large $\ell$. 
Detweiler et. al.~\cite{detweiler:03a} showed that the remaining terms scale
as $\ell^{-4}$ for large $\ell$. They find the functional form of the terms to
be
\begin{equation}
    \frac{E \mathcal{P}_{3/2}}{(2\ell-3)(2\ell-1)(2\ell+3)(2\ell+5)}
    \text{,}\label{eqn:Eterm}
\end{equation}
where $\mathcal{P}_{3/2} = 36\sqrt2$.
I fit a function of this form to the tail end of a plot of the
multipole coefficients to find the coefficient $E$ in Eq.~\eqref{eqn:Eterm}.
Extrapolating to $\ell \rightarrow \infty$ I find that the truncation error
is
\begin{align}
    \varepsilon &= \sum_{\ell = \ell_{\text{max}}}^\infty
                [\text{Eq.~\eqref{eqn:Eterm}}] \label{eqn:Eterm-sum}\\
             &= \frac{12\sqrt2 E \ell_{\text{max}}}
                     {(2\ell_{\text{max}}+3)(2\ell_{\text{max}}+1)
		      (2\ell_{\text{max}}-1)(2\ell_{\text{max}}-3)}
    \text{,}
\end{align}
where $\ell_{\text{max}}$ is the value at which I cut the summation short. 		  

A second source of error lies in the numerical calculation of the retarded
solution to the wave equation. This error depends on the step size $h$ used to
evolve the field forward in time. For a numerical scheme of a given
convergence order, I can estimate this discretization error by extrapolating
from simulations using different step sizes down to $h = 0$. 
This is what was done in the graphs shown in
Sec.~\ref{sec:convergence-tests-particle}.

\begin{figure}
    \includegraphics{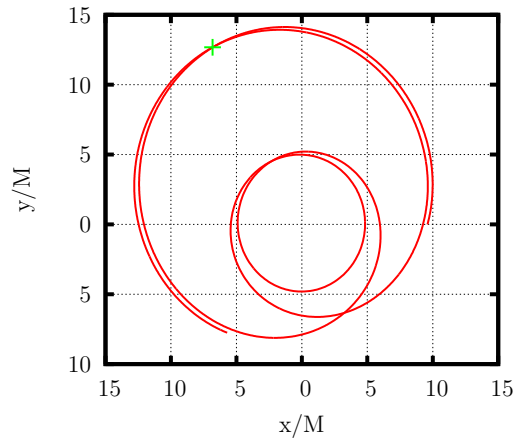}
    \caption{Trajectory of a particle with $p=7.2$, $e=0.5$. The cross-hair
    indicates the point where the data for
    Fig.~\ref{fig:multipole-coeffs-eccentric-orbit-plusminus} was extracted.}
    \label{fig:eccentric-orbit-trajectory}
\end{figure}

I display results for the mildly eccentric orbit shown in
Fig.~\ref{fig:eccentric-orbit-trajectory} with data extracted at $t = 500\,M$,
that is at the instant shown in
Fig.~\ref{fig:multipole-coeffs-eccentric-orbit-plusminus}. At this moment, the
multipole coefficients of $\real(F^R_{(+)})$ decay as expected, but \eg\ the
$\imag(F^R_{(+)})$ component decays faster with $\ell$ for the range of modes
$0\le \ell \le 13$
modes that were calculated. I choose an orbit of low eccentricity as
high eccentricity causes 
the field values to be plagued by high frequency noise, as discussed
in~\paperII. This makes it impossible to
reliably estimate the discretization error for these orbits.

Table~\ref{tab:numerical-errors} lists typical values for the errors
discussed above.
\begin{table}
    \begin{ruledtabular}
    \begin{tabular}{p{4.5cm}c}
	error estimation & mildly eccentric orbit \\
	\hline
	relative truncation error in $\frac{M^2}{q^2} \real(F^R_{(+)})$
	% gnuplot -persist <<"HERE"
	% E(x)=E*36*sqrt(2)/(2*x-3.)/(2*x-1.)/(2*x+3.)/(2*x+5.)
	% fit [8:] E(x) "<gawk '$1 == 500' eccentric-regul.out" using 6:10 via E
	% # Final set of parameters            Asymptotic Standard Error
	% # E               = -0.000897708     +/- 3.448e-05    (3.841%)
	% plot "<gawk '$1 == 500' eccentric-regul.out" using 6:(log10(abs($7))), "" using 6:(log10(abs($8))), "" using 6:(log10(abs($9))), "" using 6:(log10(abs($10))), log10(abs(E(x)))
	% el_max=13
	% print 12*sqrt(2)*E*el_max/(2*el_max+3)/(2*el_max+1)/(2*el_max-1)/(2*el_max-3)
	% # -4.39890903194733e-07
	% HERE
	% gawk '$1 == 500 {F+=$10} END {print F}' eccentric-regul.out
	% # 0.00195841
	& $2\times10^{-4}$ \\
	relative discretization error in $\frac{M^2}{q} \psi$
	% bash regress_field.sh  128,64,32 psi-regression --p 7.2 --e 0.5 --el 2 --em 2 --Delta_t 0.25 --t_max 600
	% psi = 3
	% delta = 5e-8
	& $\approx10^{-7}$ \\
    \end{tabular}
    \end{ruledtabular}
    \caption{Estimated values for the various errors in the components of the
    self-force as described in the text.
    I show the truncation and discretization errors for the mildly eccentric
    orbit ($p = 7.2$, $e = 0.5$). 
    The truncation error is calculated using a plot similar to the one shown
    in 
    Fig.~\ref{fig:multipole-coeffs-eccentric-orbit-plusminus}.
    The
    discretization error is estimated using a plot similar to that in
    Fig.~\ref{fig:sourced-regression} for the $\ell = 2$, $m = 2$
    mode.}
    \label{tab:numerical-errors}
\end{table}

\section{Sample results\label{sec:sample-results}}
In this section I describe some results obtained from my numerical calculation.
\subsection{Mildly eccentric orbit\label{sec:mildly-eccentric-orbit}}
\begin{figure}
    \includegraphics{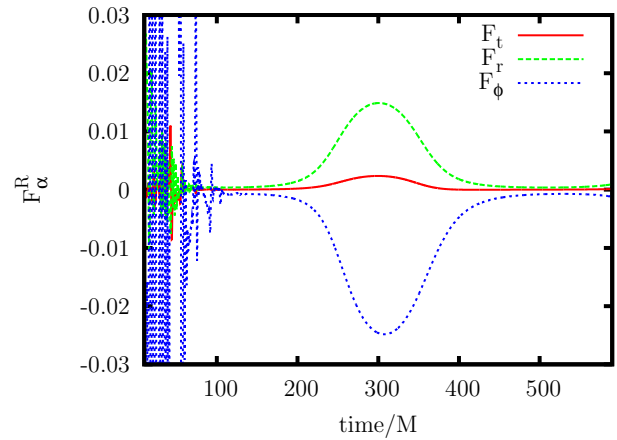}
    \caption{Regularized dimensionless self-force $\frac{M^2}{q^2} F_t$,
    $\frac{M^2}{q^2} F_r$ and $\frac{M}{q^2} F_\phi$
    on a particle on an eccentric orbit with
    $p = 7.2$, $e = 0.5$. }
    \label{fig:regularized-force}
\end{figure}

I choose a particle 
on an eccentric orbit with $p = 7.2$, $e = 0.5$ which
starts at $r = pM/(1-e^2)$, halfway between periastron and apastron.
The field is evolved for $600\,M$ with a uniform resolution of 512 grid points
per $M$, both in the $t$ and $r^*$ directions, for all values of  $\ell$.
Multipole coefficients
for $1 \le \ell \le 15$ are calculated and used to reconstruct the regularized
self-force $F_\alpha$ along the geodesic. 
Figure~\ref{fig:regularized-force} shows the result of the calculation. 
For the choice of parameters used to calculate the force
shown in Fig.~\ref{fig:regularized-force}, the error bars corresponding to the
truncation error Eq.~\eqref{eqn:Eterm-sum} 
(which are already much larger than than the discretization
error)
would be of the order of the line thickness and have not been drawn. 

Already for this small eccentricity, I see that the self-force is most important
when the particle is closest to the black hole (ie.\ for $200\,M \lesssim t
\lesssim 400\,M$). The
self-force acting on the particle is very small once the particle has moved
away to $r \approx 15\,M$. 

\subsection{Zoom-whirl orbit\label{sec:zoom-whirl-orbit}}
\begin{figure}
    \includegraphics{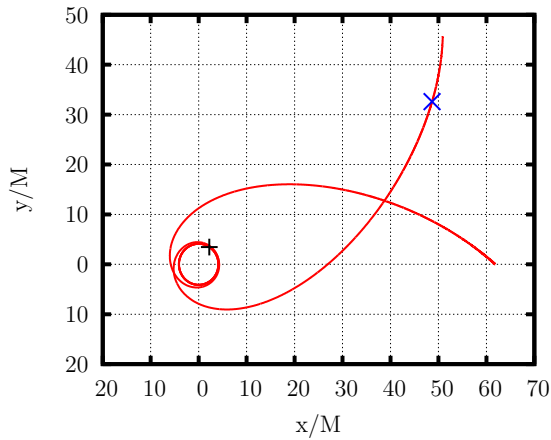}
    \caption{Trajectory of a particle on a zoom-whirl orbit with $p =
    7.8001$, $e=0.9$. The cross-hairs indicate the positions where the data
    shown in Fig.~\ref{fig:zoom-whirl-orbit-multipole-coefficients-whirling}
    and~\ref{fig:zoom-whirl-orbit-multipole-coefficients-zooming} was
    extracted.}
    \label{fig:zoom-whirl-orbit-trajectory}
\end{figure}

\begin{figure}
    \includegraphics{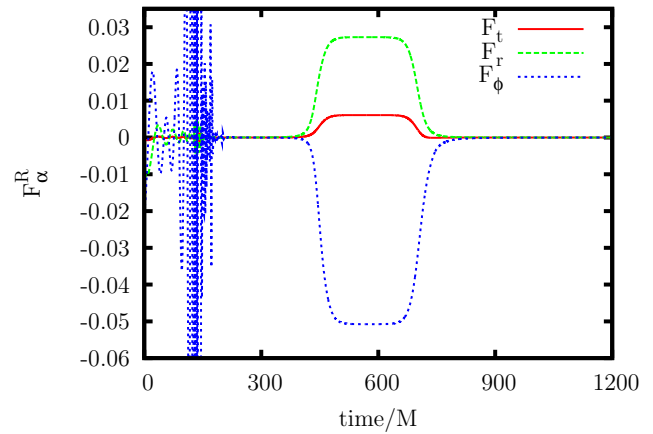}
    \caption{Self-force acting on a particle. Shown is the dimensionless
    self-force $\frac{M^2}{q^2} F_t$,
    $\frac{M^2}{q^2} F_r$ and $\frac{M}{q^2} F_\phi$ on a zoom-whirl
    orbit with $p = 7.8001$, $e=0.9$.  
    No error bars showing an estimate error are shown, as the errors shown are
    to small to show up on the graph.  Notice that
    the self-force is essentially zero during the zoom phase $900\,M \lesssim
    t \lesssim 1200\,M$ and reaches a constant value very quickly after the
    particle enters into the whirl phase.}
    \label{fig:zoom-whirl-orbit-force}
\end{figure}

Particles on highly eccentric orbits are of most interest as sources of
gravitational
radiation. For nearly parabolic orbits with $e \lesssim 1$ and $p \gtrsim 6 +
2e$, a particle revolves around the black hole a number of times,
moving on a nearly circular trajectory close to the event horizon (``whirl
phase''), before 
moving away from the black hole (``zoom phase''). During the whirl phase
the particle is in the strong field region of the spacetime, 
emitting copious amounts of radiation.
Figures~\ref{fig:zoom-whirl-orbit-trajectory} 
and~\ref{fig:zoom-whirl-orbit-force} show
the trajectory of a particle and the force 
on such an orbit with $p = 7.8001$, $e = 0.9$ calculated using a uniform step
size of $h = 1/256\,$ throughout the range $1 \le \ell \le 15$.
Even more so than for the mildly eccentric orbit discussed in
Sec.~\ref{sec:mildly-eccentric-orbit}, the self-force (and thus the amount
of radiation produced) is much larger while the particle is close to the black
hole than when it zooms out. The force graph is very similar to that obtained
for the scalar self-force in~\paperII{}, however the overshooting behaviour
at the onset and near the end of the whirl phase is not as pronounced. 

Since the rates of change in energy $E$ and angular momentum $J$ of the
trajectory are directly related to the self-force
\begin{equation}
    \dot E = -a_t\text{,} \qquad  
    \dot J =  a_\phi\text{,}
\end{equation}
it is easy to see that the self-force shown in Fig.~\ref{fig:zoom-whirl-orbit-force}
confirms the expectation that the self-force
decreases both the energy and angular momentum of the particle while radiation is
emitted. 

\begin{figure}
    \includegraphics{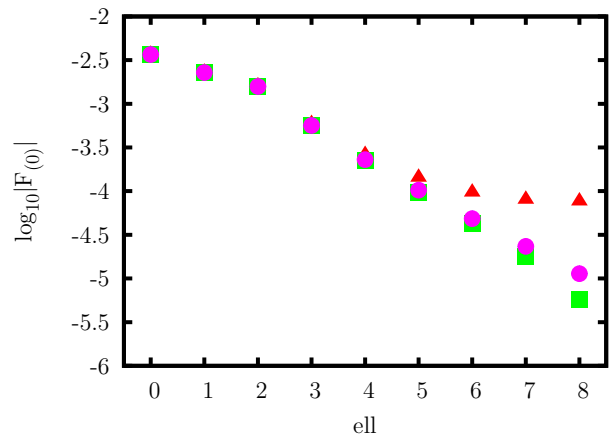}
    \caption{Multipole coefficients of $\frac{M^2}{q}\real F^R_{(0)}$ for a
    particle on a zoom-whirl orbit ($p = 7.8001$, $e = 0.9$).
    The
    coefficients are extracted at $t = 525\,M$ when the particle is deep
    within the whirl phase. Here $\dot r \approx 0$ and the behaviour of
    $F^{R}_{(\mu),\ell}$ is very close to that for a circular orbit, requiring
    very little regularization. Red triangles are used for the unregularized
    multipole coefficients $F_{(0),\ell}$, squares, diamonds and disks are
    used for the partly regularized coefficients after the removal of the
    $A_{(0)}$, $B_{(0)}$ and $D_{(0)}$ terms respectively.}
    \label{fig:zoom-whirl-orbit-multipole-coefficients-whirling}
\end{figure}

\begin{figure}
    \includegraphics{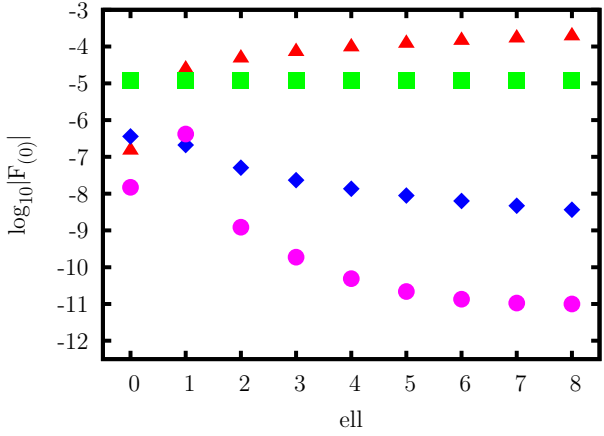}
    \caption{Multipole coefficients of $\frac{M^2}{q}\real F^R_{(0)}$ for a
    particle on a zoom-whirl orbit ($p = 7.8001$, $e = 0.9$).
    The
    coefficients are extracted at $t = 1100\,M$ when the particle is far
    away from the black hole. As $\dot r$ is non-zero, all components of the
    self-force require regularization and I see that the dependence of the
    multipole coefficients on $\ell$ is as predicted by
    Eq.~\ref{eqn:mode-by-mode-regularization}. After the removal of the regularization
    parameters $A_{(\mu)(\nu)}$, $B_{(\mu)(\nu)}$, and $D_{(\mu)(\nu)}$
    the remainder is
    proportional to $\ell^0$, $\ell^{-2}$ and $\ell^{-4}$ respectively.}
    \label{fig:zoom-whirl-orbit-multipole-coefficients-zooming}
\end{figure}

In
Fig.~\ref{fig:zoom-whirl-orbit-multipole-coefficients-whirling} 
and
Fig.~\ref{fig:zoom-whirl-orbit-multipole-coefficients-zooming} 
I show
plots of $F^\ell_{(0)}$ constructed from $F^\ell_{(\mu)(\nu)}$
after the removal of the  
$A_{(\mu)(\nu)}$, $B_{(\mu)(\nu)}$, and $D_{(\mu)(\nu)}$ terms. 

\section{Effects of the conservative self-force\label{sec:conservative-self-force}}
In this section only, I will use the subscript ``0'' to denote quantities
evaluated on the unperturbed geodesic, and no subscript to denote quantities
evaluated on the perturbed world-line. 

I
follow the literature (see \eg~\cite{pound:07}) and define
the dissipative part to be the half retarded minus half advanced force and the
conservative part to be the half retarded plus half advanced force
\begin{gather}
    F^{\text{diss}}_\alpha \define 
      \frac12 \left( F^{\text{ret}}_\alpha - F^{\text{adv}}_\alpha \right) 
      \text{,}
    \\
    F^{\text{cons}}_\alpha \define 
      \frac12 \left( F^{\text{ret}}_\alpha + F^{\text{adv}}_\alpha \right)
      \text{.} 
    \label{eqn:self-force-parts}
\end{gather}
The conservative force is the time reversal invariant part of the
self-force. It does not affect the radiated energy or angular momentum fluxes
$\dot E$ and $\dot J$; it
shifts the values of $E$ and $J$ away from their
geodesic values, affecting the orbital motion and the phase of the emitted waves. 

To obtain expressions for $E$ and $J$ under the influence of the self-force, 
I employ the procedure described in~\cite{diaz-rivera:04}.
I begin by writing down the normalization condition for the four velocity
\begin{equation}
    -1 = u^\alpha u_\alpha = -\frac{E^2}{f} + \frac{J^2}{r^2}
    \text{,}\label{eqn:conservation-of-u}
\end{equation}
as well as the $r$-component of the geodesic equation
\begin{equation}
    \frac{F^r}{m} = \ddot r - \frac{M}{(r-2M) r} \dot r^2
    - \frac{(r-2M)J^2}{r^4} + \frac{M E^2}{(r-2M) r} 
    \text{,}\label{eqn:geodesic-r}
\end{equation}
where $F^r = q F^r_{\ \mu} u^\mu$ 
is the radial component of the self-force. Solving 
Eq.~\eqref{eqn:conservation-of-u} and~\eqref{eqn:geodesic-r} I find
\begin{align}
    E^2 = E_0^2 - \frac{(r-2M)r}{r-3M} \frac{F^r}{m}
    \text{,}\label{eqn:E-solution}\\
    J^2 = J_0^2 - \frac{r^4}{r-3M} \frac{F^r}{m}
    \text{,}\label{eqn:J-solution}
\end{align}
where 
\begin{align}
    E_0^2 &= \dot r^2 + \frac{(r-2M)r \ddot r}{r-3M}+\frac{(r-2M)^2}{(r-3M)r}\text{,}\\
    J_0^2 &= \frac{r^4 \ddot r}{r-3M} + \frac{M r^2}{r-3M} \text{.}
\end{align}
I stress that $E_0$ and $J_0$ are not the geodesic \emph{values} for energy and
angular momentum. They are of the correct form but are evaluated using the
\emph{accelerated} values for $r$, $\dot r$ and $\ddot r$ (instead of the
geodesic values $r_0$, $\dot r_0$, etc.). 

For small perturbing force of order $\varepsilon$ I expand
Eqs.~\eqref{eqn:E-solution} and~\eqref{eqn:J-solution} in terms of the
perturbation strength and find
\begin{align}
    E = E_0 + \Delta E \approx E_0 - \varepsilon \frac{(r-2M)r}{2 (r-3M) E_0}
    \frac{F^r}{m} + \orderof{\varepsilon^2}\text{,}\label{eqn:E-expanded-solution}\\ 
    J = J_0 + \Delta J \approx J_0 - \varepsilon \frac{r^4}{2(r-3M) J_0} 
    \frac{F^r}{m} + \orderof{\varepsilon^2}\text{,}\label{eqn:J-expanded-solution}
\end{align}
where $F^r$ is evaluated with the help of the unperturbed four velocity
$u_0^\alpha = [E_0/f, \dot r_0, 0, J_0/r_0^2]$. The fractional changes $\Delta
E/E_0$ and $\Delta J/J_0$ are given by
\begin{align}
    \Delta E/E_0  = - \varepsilon \frac{(r-2M)r}{2 (r-3M) E_0^2}
    \frac{F^r}{m} +
    \orderof{\varepsilon^2}\text{,}\label{eqn:E-expanded-relative}\\ 
    \Delta J/J_0  = - \varepsilon \frac{r^4}{2(r-3M) J_0^2} 
    \frac{F^r}{m} +
    \orderof{\varepsilon^2}\text{.}\label{eqn:J-expanded-relative}
\end{align}

Once the perturbations in $E$ and $J$ are known, I calculate the change in
the angular frequency
\begin{equation}
    \Omega \define \odiff{\varphi}{t} =
    \frac{r-2M}{r^3}\frac{J}{E}\text{.}\label{eqn:def-Omega}
\end{equation}
For small perturbing force I expand in powers of the perturbation strength
\begin{multline}
    \Omega = \frac{r_0-2M}{r_0^3}\frac{J_0}{E_0} 
    \biggl[1 - \varepsilon \biggl( 
      \frac{r^4}{2(r-3M) J_0^2} 
    \\
    - \frac{(r-2M)r}{2 (r-3M) E_0^2}
    \biggr)
    \frac{F^r}{m} 
    \biggr]
    + \orderof{\varepsilon^2}
    \text{.}\label{eqn:Omega-expanded-solution}
\end{multline}
The relative change $\Delta\Omega/\Omega_0$ is given by
\begin{multline}
    \Delta\Omega/\Omega_0 = - \varepsilon \biggl( 
      \frac{r^4}{2(r-3M) J_0^2} 
    - \frac{(r-2M)r}{2 (r-3M) E_0^2}
    \biggr)
    \frac{F^r}{m} 
    \\
    + \orderof{\varepsilon^2}
    \text{.}\label{eqn:Omega-expanded-relative}
\end{multline}

\subsection{Circular
orbits}\label{sec:conservative-self-force-on-circular-orbits}
The effect of the conservative self-force is
most clearly observed for circular orbits, where the unperturbed angular
frequency $\Omega$ as well as the shift due to the perturbation are constant
in time. 

For a particle in circular motion the self-force is constant in time and it
turns out that the radial component is entirely conservative whereas the 
$t$ and $\phi$ components are entirely dissipative. 
For circular orbits, the unperturbed values of $E$ and $J$ are given by 
\begin{align}
    E_0 &= \frac{r_0-2M}{\sqrt{r_0(r_0-3M)}} \text{,}\\
    J_0 &= r_0 \sqrt{\frac{M}{r_0-3M}} \text{,}
\end{align}
and substituting these into Eq.~\eqref{eqn:Omega-expanded-solution} I find
\begin{equation}
    \Omega = \sqrt{\frac{M}{r_0^3}} - \frac{(r_0-3M)}{2mM} \sqrt{\frac{M}{r_0}} F_r +
    \orderof{\varepsilon^2}
    \text{,}
\end{equation}
where the first term is just the angular frequency for an unperturbed geodesic
at radius $r_0$. The fractional change $\Delta\Omega/\Omega_0$ is then
\begin{equation}
    \frac{\Delta\Omega}{\Omega_0} = - \frac{(r_0-3M) r_0}{2mM} F_r +
    \orderof{\varepsilon^2}
    \text{.}
\end{equation}
Similarly the fractional changes in $E$ and $J$: $\Delta E/E_0$ and
$\Delta J/J_0$ are given by
\begin{align}
    \Delta E/E_0 &= -\frac{r_0}{2m} F_r \text{,}\\
    \Delta J/J_0 &= -\frac{(r_0-2M) r_0}{2mM} F_r
    \text{.} 
\end{align}
\begin{figure}
    \includegraphics[width=8.6cm]{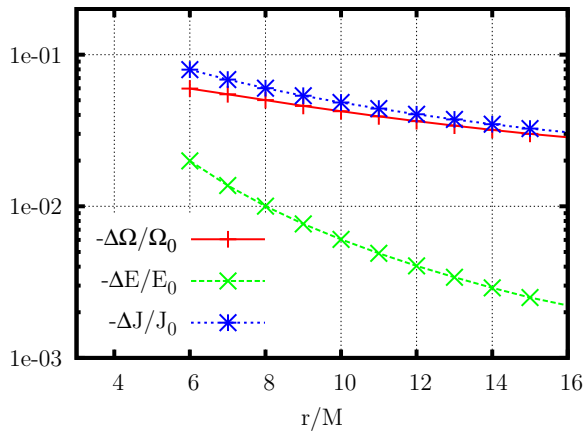}
    \caption{Fractional change $\Delta\Omega/\Omega_0$ induced by the presence
    of the conservative self-force. The effect of the self-force is to move the
    radius of the orbit outward, decreasing its angular frequency.
    \label{fig:fractional-change-in-omega-circular}}
\end{figure}
Figure~\ref{fig:fractional-change-in-omega-circular} shows the fractional
change in $\Omega_0$, $E$ and $J$ as a function of the orbit's radius $r_0$. 

\subsection{Eccentric orbits}
For eccentric orbits the self-force is no longer constant in time and I have
to numerically calculate both the retarded and the advanced self-force in
order to construct the conservative self-force. I find the advanced force by
running the simulation backwards in time. That is I start the evolution on the
very last time slice and evolve backwards in time until I reach the slice
corresponding to $t = 0$. I reverse the boundary condition at the event
horizon to be be outgoing radiation only $(\partial_t + \partial_{r^*}) \psi
= 0$ and adjust the outer boundary so as to simulate only the backwards domain
of dependence of the initial slice. I do not change the trajectory of the
particle. I do not change the regularization parameters, since they depend
only on the local behaviour of the field and are insensitive to the boundary
conditions far away.

\subsubsection{Conservative force on zoom-whirl orbits} 
\begin{figure}
    \includegraphics{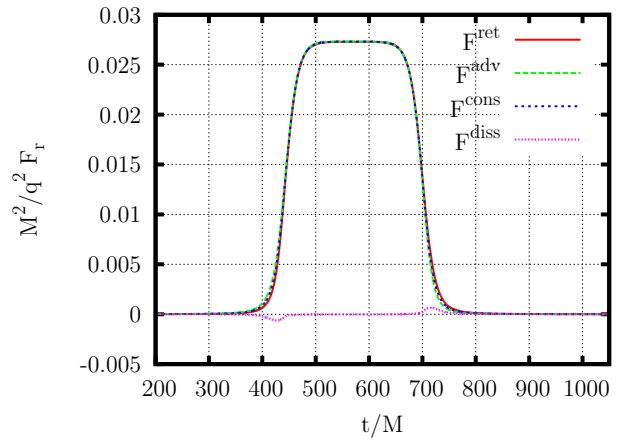}
    \caption{$r$
    component of the dimensionless self-force acting on a particle on a
    zoom-whirl orbit ($p=7.8001$, $e=0.9$) around a Schwarzschild black hole.
    Shown are the retarded (solid, red), advanced (dashed, green),
    conservative (dotted, blue) and dissipative (finely dotted, pink) force
    acting on the particle.}\label{fig:zoom-whirl-orbit-cons-diss-force-r}
\end{figure}

\begin{figure}
    \includegraphics{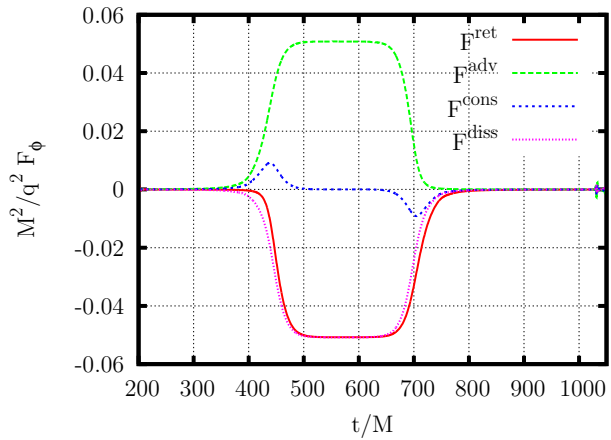}
    \caption{$\varphi$
    component of the dimensionless self-force acting on a particle on a
    zoom-whirl orbit ($p=7.8001$, $e=0.9$) around a Schwarzschild black hole.
    Shown are the retarded (solid, red), advanced (dashed, green),
    conservative (dotted, blue) and dissipative (finely dotted, pink) force
    acting on the particle.}\label{fig:zoom-whirl-orbit-cons-diss-force-phi}
\end{figure}

I calculate the
conservative self-force on a zoom-whirl orbit with $p=7.8001$, $e = 0.9$.
Figs.~\ref{fig:zoom-whirl-orbit-cons-diss-force-r}
and~\ref{fig:zoom-whirl-orbit-cons-diss-force-phi}
display the breakdown of the self-force into retarded and
advanced, and conservative and dissipative parts for a particle on a
zoom-whirl orbit. In both plots the force is very weak when the particle is in
the zoom phase $t \lesssim 400\,M$ or $t \gtrsim 800 \, M$ and nearly
constant while the particle is in the whirl phase $400\,M \lesssim t \lesssim
800\,M$.  Inspection of the behaviour of the $r$ component reveals that it is
almost exclusively conservative, with only a tiny dissipative effect when the
particle enters or leaves the whirl phase. This result is consistent with the
observation that the particle moves on a nearly circular trajectory while in
the whirl phase, for which the radial component is precisely conservative.
Similarly the $\phi$ component is almost entirely dissipative, with only a
small conservative contribution when the particle enters or leaves the whirl
phase, its maximum coinciding with that of $\ddot r$ (not shown on the graph). 

\begin{figure}
    \includegraphics{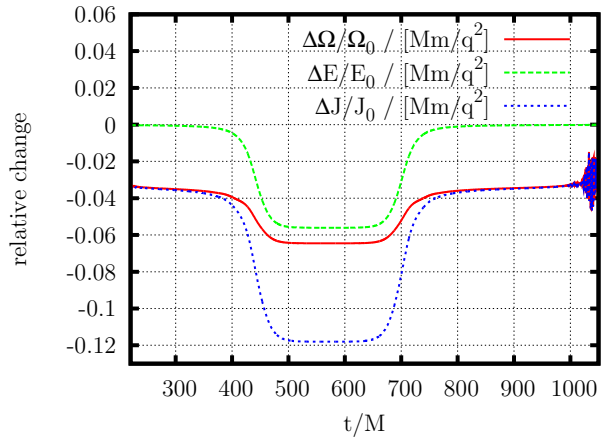}
    \caption{Relative change in $\Omega$, $E$, $J$ for a particle on a
    zoom-whirl orbit due to the conservative electromagnetic 
    self-force.}\label{fig:zoom-whirl-orbit-fractional-change-in-omega}
\end{figure}

I calculate the relative changes in $E$, $J$ and $\Omega$ under the influence
of the self-force using Eqs.~\eqref{eqn:E-expanded-relative},
\eqref{eqn:J-expanded-relative}, \eqref{eqn:Omega-expanded-relative}. 
Fig.~\ref{fig:zoom-whirl-orbit-fractional-change-in-omega} 
displays the
relative changes $\Delta E/E_0$, $\Delta J/J_0$ and $\Delta\Omega/\Omega_0$
for a particle on a zoom whirl orbit $p=7.8001$, $e=0.9$. 
The change in $E$, $J$ and $\Omega$ is strongest in the whirl phase when $r
\approx 4.1M$. It is 
consistent with the shift experienced by a particle on a circular orbit at
$4.1M$. 

\subsection{Effects on the innermost stable orbit}
In the gravitational case, considerable work has been done to identify gauge
invariant effects of the self-force~\cite{Detweiler:2008ft, Sago:2008id}. The
electromagnetic
self-force is not subject to the same ambiguity thus it can help shed light on
the gravitational case as well by providing a clear distinction between
kinetic and dynamic effects.
In this section I calculate the effect of
the conservative self-foce on the location of the innermost stable circular
orbit around a Schwarzschild black hole. Such a calculation was first
performed for the scalar self-force by~\cite{diaz-rivera:04}, where a highly
accurate frequency domain numerical scheme was used. Recently~\cite{barack:09,
barack:2011ed}
have extended this calculation to gravity, using their time domain code to
perform the intergration of the wave equation. Since the code presented in
this paper is in the time domain as well, it is closest in spirit
to~\cite{barack:09}.

\section{Retardation of the
self-force}\label{sec:retardation-of-self-force} 
For scalar perturbation in a weak gravitational field
Poisson~\cite{pfenning:00} showed that the self-force is
delayed with respect to the particle motion by the light travel time from the
particle to the central body and back to the particle again. 
In a spacetime where the central body is compact the treatment
of~\cite{pfenning:00} is no longer directly applicable, but I
still expect some retardation in the self-force when compared to the
particle's motion. 
To study this effect, I calculate the self-force on an eccentric orbit with
$p=78$, $e=0.9$; ten times larger than the zoom-whirl orbit discussed earlier.
The large orbit was chosen so as to be able to clearly see any possible
retardation which might not be visible if the particle's orbit is deep within
the strong field region close to the black hole. 
\begin{figure}
    \includegraphics{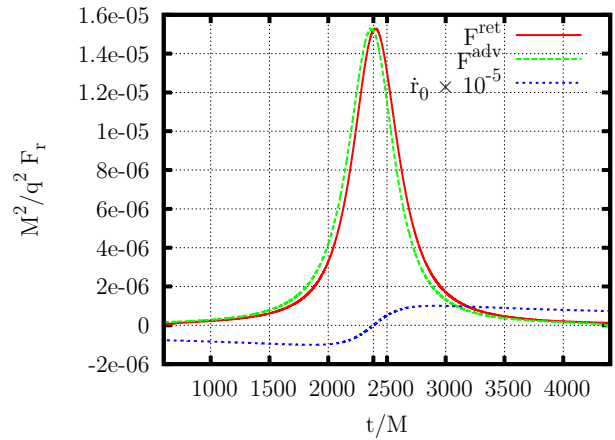}
    \caption{$r$ component of the dimensionless self-force acting on a
    particle on an orbit with $p=78$, $e=0.9$. Shown are the retarded and
    advanced forces as well as $\dot r$. The vertical line at $t\approx
    2383\,M$
    marks the time of closest approach to the black hole.}\label{fig:conservative-force-r}
\end{figure}
\begin{figure}
    \includegraphics{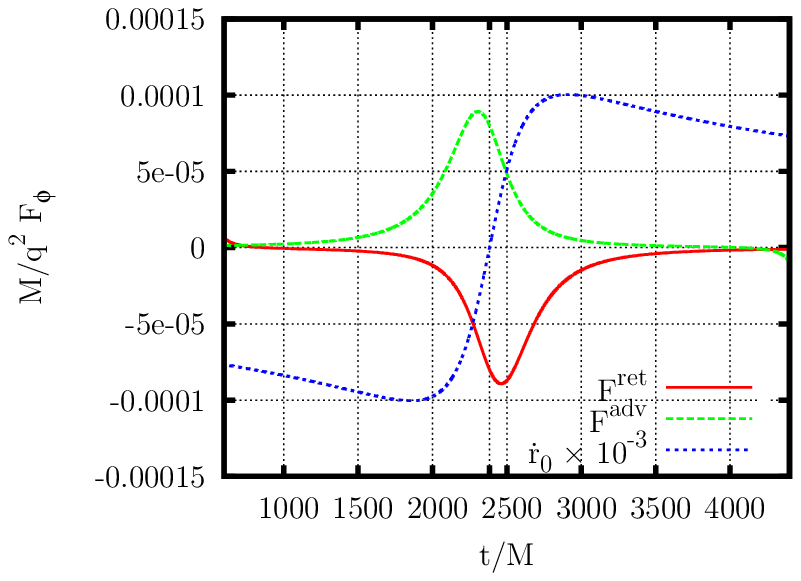}
    \caption{$\phi$ component of the dimensionless self-force acting on a
    particle on an orbit with $p=78$, $e=0.9$. Shown are the retarded and
    advanced forces as well as  $\dot r$. The vertical line at
    $t\approx2383\,M$
    marks the time of closest approach to the black hole.}\label{fig:conservative-force-phi}
\end{figure}
Figures~\ref{fig:conservative-force-r}
and~\ref{fig:conservative-force-phi} display plots of the $r$ and
$\phi$ components of the self-force acting on the particle close to
periastron. Shown are the retarded and advanced forces as well as the
particle's radial velocity $\dot r$. Without considering retardation I 
expect the 
self-force to be strongest
when the particle is closest to the black hole, when $\dot r = 0$, as evident
in Fig.~\ref{fig:zoom-whirl-orbit-cons-diss-force-r}. 
Clearly for the $r$ component displayed in Fig.~\ref{fig:conservative-force-r}
the retarded and advanced forces both peak at a time very close to the zero
crossing of $\dot r$, suggesting very little time delay in the $r$ component
of the self-force. In Fig.~\ref{fig:conservative-force-phi} on the other hand
the retarded and advanced $\phi$-component of the self-force peaks away from
the time of closest approach
$t_{\text{min}}$. Inspection of the graph shows that the delay (advance)
between the time of closest approach and the peak in the retarded (advanced)
force is compatible with a delay of $\Delta t_{\text{min}} \approx 2
(r_{\text{min}} - 3.0\,M)\approx 74\,M$. Using a delay of $\Delta t \approx
2[r_0(t)-3.0\,M]$ and plotting $F^{\text{ret}}_\varphi(t+\Delta t)$ and
$-F^{\text{adv}}_\varphi(t-\Delta t)$ versus $t$ both curves visually lie on top of
each other and the maximum is located at $t_{\text{min}}$ as shown in
Fig.~\ref{fig:conservative-force-phi-shifted} below.
This suggests that the self-force is in large parts due to radiation that
travels into the strong field region close to the black hole and is
scattered back to the particle. The time delay can then be loosely interpreted
as the time it takes the signal to travel to the light ring around the black
hole and back to the particle. This interpretation is loose for two reasons:
First $r^*$ and not $r$ is associated with the light travel time.
Using $r^*$, however, does not lead to a better overlap of the curves once a
suitable constant offset chosen. Second, for the zoom-whirl orbit
shown in Fig.~\ref{fig:zoom-whirl-orbit-force} the (shallow) maximum in the
self-force is offset by only $\Delta t \approx 2[r_0(t)-1.0\,M]$ which leads to a
reasonable overlap of the two curves. Interestingly using $r^*$ instead of $r$
yields a worse overlap. For very large orbits $p = 780$, $e=0.9$ it
is impossible to read off the small constant offset to the dominant $2r_0(t)$
contribution. 

\section{Weak field limit}\label{sec:weak-field-limit}
\begin{figure}
    \includegraphics{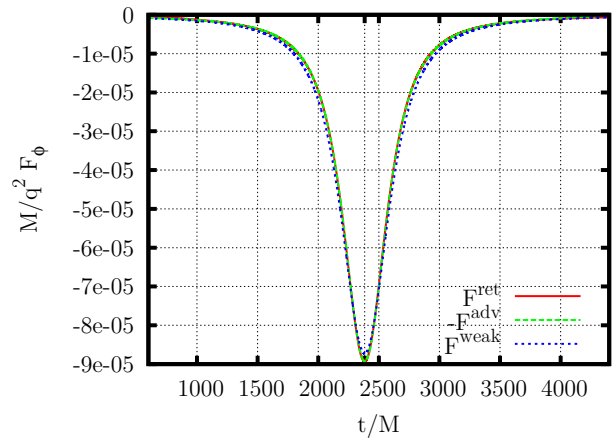}
    \caption{$\phi$ component of the retarded (solid red line) and (dashed
    green line) negative advanced self-forces acting on a particle with $p=78$,
    $e=0.9$. The forces have been shifted by $\Delta t \approx
    2[r_0(t)-3.0\,M]$. Also shown is the self-force calculated using the weak
    field expression Eq.~\ref{eqn:weak-field-self-force} (blue dotted line).}
    \label{fig:conservative-force-phi-shifted}
\end{figure}

As a last application I use my code to compare the numerical self-force in the weak
field region to the self-force calculated using the weak field expression
\begin{equation}
    \vec f_{\text{self}} = 
    \lambda_c \frac{q^2}{m} \frac{M}{r^3} \hat{\vec r}
      + \lambda_{rr} \frac23 \frac{q^2}{m} \odiff{\vec g}{t}
      \text{,}\qquad
      \vec g = -\frac{M}{r^2} \hat{\vec r}
    \text{,}\label{eqn:weak-field-self-force}
\end{equation}
of~\cite{pound:07,dewitt:64}.
I calculate the self-force for a particle on an eccentric
orbits with $e=0.9$ and $p=78$ or $p=780$.
Fig.~\ref{fig:conservative-force-phi-shifted} 
shows the retarded and
(negative) advanced
forces shifted by $\Delta t \approx 2[r_0(t)-3.0\,M]$ as well as the analytic
force calculated using Eq.~\eqref{eqn:weak-field-self-force}. At this distance
there are still some differences between the (shifted) retarded field
and the weak field expression. One reason for this lies in the choice of a
suitable $r$ coordinate to correspond to the $r$ coordinate in the weak field
expression. In this work I use the areal Schwarzschild $r$, but the isotropic
coordinate $\bar r = \frac{r-M}{2}+\frac{\sqrt{r\,(r-2*M)}}{2}$ or even 
the tortoise $r^*$ could be used as well. Neither one
yields a good agreement between the two curves. 

\begin{figure}
    \includegraphics{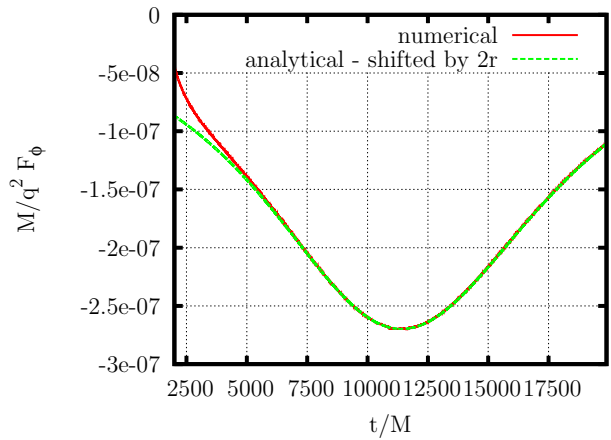}
    \caption{$\phi$ component of the retarded self-force acting on a particle on an orbit
    with $p=780$, $e=0.9$ close to periastron. Shown are the numerical (solid,
    red) and shifted analytical (dashed, green) forces. The agreement between numerical and analytical
    calculation is excellent, the discrepancy for $t \lesssim 7500\,M$ is due
    to initial data contamination.\label{fig:weak-field-force-phi}}
\end{figure}

For $p=780$ using a shift of $\Delta t
= 2r_0(t)$ the agreement between numerical data and analytic expression is
excellent as is evident in Fig.~\ref{fig:weak-field-force-phi}. At this
distance $r$, $\bar r$ and $r^*$ are indistinguishable. 

\section{Conclusions}\label{sec:conclusions}
I calculated the self-force acting 
a on an electromagnetic point charge in orbit around a Schwarzschild black hole.
To do so I calculated the regularization
parameters $A$,$B$, and $D$ in section~\ref{sec:regularization-parameters} and
implemented a second order accurate numerical scheme in
section~\ref{sec:numerical-method}. 

I find the behaviour of the
electromagnetic self-force to be similar but not identical to that of the
scalar self-force. In both cases the self-force is strongest when the particle is
closest to the black hole. Further, during the whirl phase of a zoom-whirl
orbit with its nearly
constant radius, the self-force is very close to that of a particle in circular orbit at
this radius. On the other hand, the overshooting effect upon entering the
whirl phase which was observed in the scalar case is much weaker in the
electromagnetic case. 

I calculated the effects of the conservative self-force on circular orbits,
where it reduces the angular frequency and thus affects the phasing of the
observed waves. I find this effect to be much stronger in the
electromagnetic case than in the scalar case discussed
in~\cite{diaz-rivera:04}. In particular during the nearly
circular whirl phase of a zoom-whirl orbit I find that $\Omega$ decreases by 
$\approx 0.06 \frac{q^2}{Mm}$. Due to the smallness of the ratio
$\frac{q^2}{Mm}$ this change is tiny for one orbit, however since it
accumulates over the inspiral, its effect on the total phase shift during the
full inspiral can be of order unity. 
 This statement 
is not directly transferable to the gravitational case since the radius $r_0$
of the orbit is not a gauge invariant quantity. Therefore I cannot
distinguish between changes in $\Omega$ due to effects of the self-force and
due to gauge choices. To obtain a meaningful measure of the effect of the
gravitational 
self-force I need to compare two gauge invariant quantities, \eg\ $\Omega$ and
the gauge invariant $u_t$ of~\cite{detweiler:08}. 

I investigated the retardation of the self-force with respect to the motion of
the particle. I found that the retardation is very weak for the $r$ component
of the force and strong in the $t$ and $\varphi$ components, which are linked
to radiated energy and angular momentum. In the later cases the retardation is
compatible with a delay of $\Delta t \approx 2 (r_0(t) - R_{\text{delay}})$, where
$R_{\text{delay}}$ is a constant depending on the particle's orbit. 

\begin{acknowledgments}
    I thank Eric Poisson and Steve Detweiler for useful discussions and
    suggestions. I gratefully acknowledge support by the Natural Sciences and
    Engineering Council of Canada.
    This work was supported in part by NSF grants PHY-0903973 and PHY-0904015.
    This work was made possible by the
    facilities of the Shared Hierarchical
    Academic Research Computing Network (SHARCNET:\url{www.sharcnet.ca}) as
    well as the e FoRCE cluster at Georgia Tech.
\end{acknowledgments}

\appendix 
\section{Translation tables\label{sec:translation-tables}}
I require coupling coefficients to translate
between the tensor harmonic modes of the Faraday tensor 
and the scalar harmonic modes of the tetrad components of the Faraday tensor.
%in terms of the complex tetrad introduced in Eqs.~\eqref{eqn:tetrad-def-0} --
%\eqref{eqn:tetrad-def-pm}.

As a first step, I reconstruct the Faraday tensor modes from the numerical
variables. For the even mode auxiliary fields $\psi$, $\chi$ and $\xi$ 
this reconstruction can be done algebraically while the odd sector requires a
numerical differentiation of the numerical variable $\tilde v$. The
reconstruction relations were already displayed in
Eqs.~\eqref{eqn:faraday-from-auxiliary-fields-begin} --
\eqref{eqn:faraday-from-auxiliary-fields-end}, which involves both the even
and odd modes.

In terms of the vector
potential the Faraday tensor modes are reconstructed using the defining 
equation
Eq.~\eqref{eqn:faraday-from-vector-potential}. In this case, the
reconstruction of the Faraday tensor reads
\begin{subequations}
\begin{align}
    F_{tr} &= \sum_{\ell,m} (A^{\ell m}_{r,t} - A^{\ell m}_{t,r}) \, Y^{\ell
    m}\text{,}\label{eqn:faraday-from-vector-harmonics-begin} \\
    F_{tA} &= \sum_{\ell,m} [ (v^{\ell m}_{,t} - A^{\ell m}_t) \, Z^{\ell m}_A
    + \tilde v^{\ell m}_{,t} \, X^{\ell m}_A]\text{,}\\
    F_{rA} &= \sum_{\ell,m} [ (v^{\ell m}_{,r} - A^{\ell m}_r) \, Z^{\ell m}_A + \tilde v^{\ell m}_{,r}
    \, X^{\ell m}_A]\text{,}\\
    F_{\theta\varphi} &= 
    \sum_{\ell,m} \tilde v_{\ell m} \, (X^{\ell m}_{\varphi,\theta} - X^{\ell m}_{\theta,\varphi})
    \text{.}\label{eqn:faraday-from-vector-harmonics-end}
\end{align}
\end{subequations}
Clearly both the expansion Eqs.~\eqref{eqn:faraday-from-auxiliary-fields-begin} --
\eqref{eqn:faraday-from-auxiliary-fields-end} and the one in
Eqs.~\eqref{eqn:faraday-from-vector-harmonics-begin} --
\eqref{eqn:faraday-from-vector-harmonics-end} are of the same form and it is
only necessary to obtain one set of translation coefficients to handle both
the calculation using $\psi$, $\chi$ and $\xi$ in the main text and the one
using the vector potential that will be presented in
appendix~\ref{sec:vector-pot-calculation}. 

The tetrad components $F_{(\mu)(\nu)}$ are decomposed in terms of scalar
spherical harmonics 
\begin{equation}
    F_{(\mu)(\nu)} = \sum_{\ell,m} F^{\ell m}_{(\mu)(\nu)} Y_{\ell m}\text{,}
\end{equation}
where each mode is given by
\begin{equation}
    F^{\ell m}_{(\mu)(\nu)} = \int F_{(\mu)(\nu)} \conj{Y}_{\ell m} \,\d\Omega
    \text{.}\label{eqn:scalar-spherical-decomposition}
\end{equation}
To obtain expressions for the coupling coefficients I substitute
$F_{(\mu)(\nu)} = F_{\alpha\beta} e^\alpha_{\ (\mu)} e^\beta_{\ (\nu)}$ into
Eq.~\eqref{eqn:scalar-spherical-decomposition} 
\begin{align}
    F^{\ell m}_{(\mu)(\nu)} &= \int \d \Omega \, F_{(\mu)(\nu)}
      \conj{Y}^{\ell m}  \nonumber \\
    &= \int \d \Omega \, \left(
      A_{\beta,\alpha}-A_{\alpha,\beta} \right) e^\alpha_{\ (\mu)} e^\beta_{\ (\nu)}
      \conj{Y}^{\ell m}  \nonumber \\
    &= \int \sum_{\ell',m'} \Bigl[ 
    \left(
      A_{b,a}-A_{a,b} \right) e^a_{\ (\mu)} e^b_{\ (\nu)}
      \conj{Y}^{\ell m}  
      \nonumber\\&\quad+
    \left(
      A_{b,A}-A_{A,b} \right) e^A_{\ (\mu)} e^b_{\ (\nu)}
      \conj{Y}^{\ell m} 
      \nonumber\\&\quad +
    \left(
      A_{B,a}-A_{a,B} \right) e^a_{\ (\mu)} e^B_{\ (\nu)}
      \conj{Y}^{\ell m} 
      \nonumber\\&\quad +
    \left(
      A_{B,A}-A_{A,B} \right) e^A_{\ (\mu)} e^B_{\ (\nu)}
      \conj{Y}^{\ell m}
      \Bigr] \d\Omega
      \nonumber\\
    &\define \sum_{\ell', m'} \left[
      C^{a b}_{(\mu)(\nu)}(\ell' m' | \ell m) \left( A^{\ell'm'}_{b,a}-A^{\ell' m'}_{a,b} \right) 
      \right.\nonumber\\&\quad\left.
      + D^a_{(\mu)(\nu)}(\ell' m' | \ell m) \left( \partial_a v^{\ell'm'}-A^{\ell' m'}_a \right)  
      \right.\nonumber\\&\quad\left.
      + E^a_{(\mu)(\nu)}(\ell' m' | \ell m) \partial_a \tilde v^{\ell' m'} 
      \right.\nonumber\\&\quad\left.
      + E_{(\mu)(\nu)}(\ell' m' | \ell m) \tilde v^{\ell' m'}
      \right]
      \text{,}\label{eqn:translation-table}
\end{align}
which defines the coupling coefficients. 
It is often possible to express these coupling coefficients in terms of
linear combinations of the coupling coefficients derived in \paperI{} for the
scalar field. 

To simplify the notation of the coupling coefficients I use  
\begin{align}
    % delta
    \ccp{\ell m} &= \sqrt{\frac{(\ell+m)(\ell+m+1)}{(2\ell+1)(2\ell+3)}}
    \text{,} \\ 
    \ccm{\ell m} &= \sqrt{\frac{(\ell+m+1)(\ell-m+1)}{(2\ell+1)(2\ell+3)}}
    \text{,}
\end{align}
as shorthands for recurring combinations of terms. With these the
reusable scalar coupling coefficients are written as
\begin{align}
    C^r_{(+)}(\ell' m' | \ell m) &= 
    -\ccp{\ell-1, m} \sqrt f \delta_{\ell'\ell-1} \delta_{m' m-1}
    \nonumber \\ &\quad
    +\ccp{\ell, -m+1} \sqrt f \delta_{\ell'\ell+1} \delta_{m' m-1}
    \\
    C_{(+)}(\ell' m' | \ell m) &= 
    \ccp{\ell-1, m}\frac{\ell-1}{r} \delta_{\ell' \ell-1} \delta_{m' m-1} 
    \nonumber \\ &\quad
    +\ccp{\ell, -m+1} \frac{\ell+2}{r} \delta_{\ell' \ell+1} \delta_{m' m-1}
    \text{,}
    \\
    C_{(-)}(\ell' m' | \ell m) &= 
    -\ccp{\ell-1, -m} \frac{\ell-1}{r} \delta_{\ell' \ell-1} \delta_{m' m+1} 
    \nonumber \\ &\quad
    -\ccp{\ell, m+1} \frac{\ell+2}{r} \delta_{\ell' \ell+1} \delta_{m' m+1}
    \text{.}
\end{align}
Similarly it proves useful to define lower order coupling coefficients for
the odd sector, which is absent in the scalar case.
\begin{align}
    E_{(+)}(\ell' m' | \ell m) = -\frac{i}{r}\sqrt{(\ell-m+1)(\ell+m)} \delta_{\ell'\ell}\delta_{m' m-1}
    \text{,}
    \\
    E_{(-)}(\ell' m' | \ell m) = -\frac{i}{r}\sqrt{(\ell+m+1)(\ell-m)} \delta_{\ell'\ell} \delta_{m' m+1}
    \text{.}
\end{align}

In terms of the scalar coupling coefficients,
the first coefficient for the expansion of $F_{(0)(+)}$ is given by
\begin{align}
    C^{tr}_{(0)(+)}(\ell'm'|\ell m) &= \int Y^{\ell' m'} e^t_{(0)} e^r_{(+)} \conj{Y}^{\ell m}
    \,\d\Omega \nonumber\\
    &= \frac{1}{\sqrt{f}} \int Y^{\ell' m'} e^r_{(+)} \conj{Y}^{\ell m}
    \,\d\Omega \\
    &= C^r_{(+)}(\ell'm'|\ell m) / \sqrt{f}  \nonumber
    \text{,}
\end{align}
and all other combinations of $a$,
$b$ and $(\mu)$, $(\nu)$ lead to vanishing $C^{ab}_{(\mu)(\nu)}$. 
Similarly for the remaining non-vanishing coefficients for the $F_{(0)(+)}$
component
\begin{align}
    D^t_{(0)(+)}(\ell' m' | \ell m)    &= C_{(+)}(\ell' m' | \ell m)/\sqrt f
    \text{,}
    \\
    E^t_{(0)(+)}(\ell' m' | \ell m)    &= E_{(+)}(\ell' m' | \ell m)/\sqrt f
    \text{.}
\end{align}

The coupling coefficients for $F_{(+)(-)}$ contain both even and odd modes.
The first non-vanishing one is $D^r_{(+)(-)}(\ell'm'|\ell m)$, which is given
by
\begin{align}
    D^r_{(+)(-)}(\ell' m' | \ell m) &= 
    \sqrt f \ccp{\ell, -m+1} C_{(-)}(\ell' m' | \ell+1, m-1) 
    \nonumber\\&\quad
    -\sqrt f \ccp{\ell-1, m} C_{(-)}(\ell' m' | \ell-1, m-1)
    \nonumber\\&\quad
    +\sqrt f \ccp{\ell, m+1} C_{(+)}(\ell' m' | \ell+1, m+1) 
    \nonumber\\&%\quad
    -\sqrt f \ccp{\ell-1, -m} C_{(+)}(\ell' m' | \ell-1, m+1)
    \text{,}
\end{align}
while the coefficients coupling to odd modes are 
\begin{align}
    E^r_{(+)(-)}(\ell' m' | \ell m) &=
    \sqrt f \ccp{\ell, -m+1} E_{(-)}(\ell' m' | \ell+1, m-1)
    \nonumber\\&\quad
    -\sqrt f \ccp{\ell-1, m} E_{(-)}(\ell' m' | \ell-1, m-1)
    \nonumber\\&\quad
    +\sqrt f \ccp{\ell, m+1} E_{(+)}(\ell' m' | \ell+1, m+1) 
    \nonumber\\&%\quad
    -\sqrt f \ccp{\ell-1, -m} E_{(+)}(\ell' m' | \ell-1, m+1)
    \text{,}
    \\
    E_{(+)(-)}(\ell' m' | \ell m) &=
    -2 i/r^2 (\ell+1) (\ell+2)\ccm{\ell m} \delta_{\ell' \ell+1} \delta_{m'm}
    \nonumber\\&\quad
    -2 i/r^2 (\ell-1)  \ell   \ccm{\ell-1,m} \delta_{\ell' \ell-1} \delta_{m'm}
    \text{.}
\end{align}

\section{Regularization parameters\label{sec:regularization-parameters}}
I mimic the treatment in~\paperI{} and start from the covariant
expression for the singular vector potential tensor in Eq.~(464) of~\cite{poisson:04}
\begin{multline}
    \nabla_\beta A^S_\alpha(x) = 
    -\frac{q}{2r^2}U_{\alpha\beta'} u^{\beta'} \nabla_\beta r 
    -\frac{q}{2\radv^2}U_{\alpha\beta''} u^{\beta''} \nabla_\beta \radv
    \\\mbox{}
    +\frac{q}{2r}U_{\alpha\beta';\beta}u^{\beta'}
    +\frac{q}{2r} U_{\alpha\beta';\gamma'}u^{\beta'}u^{\gamma'} \nabla_\beta u
    +\frac{q}{2\radv}U_{\alpha\beta'';\beta}u^{\beta''}
    \\\mbox{}
    +\frac{q}{2\radv}U_{\alpha\beta'';\gamma''}u^{\beta''}u^{\gamma''} \nabla_\beta v
    +\frac12 q V_{\alpha\beta'}u^{\beta'} \nabla_\beta u
    \\\mbox{}
    -\frac12 q V_{\alpha\beta''}u^{\beta''} \nabla_\beta v
    -\frac12 q \int_u^v \nabla_\beta V_{\alpha\mu}\bm(x,z(\tau)\bm)
    u^\beta(\tau)\,\d\tau
    \text{.}\label{eqn:singular-vector-gradient}
\end{multline}
I have introduced a large number of symbols. $x$ is the point where the
field is evaluated, $x'$ and $x''$ are the retarded and advanced points of $x$ on
the world line $z(\tau)$.
They are connected to $x$ with unique future-directed and
past-directed null geodesics, respectively. $u(x)$ and $v(x)$ are the retarded
and advanced time functions such that $x' = z(\tau = u)$, $x'' = z(\tau = v)$.
$u^{\alpha'}$ and $u^{\alpha''}$ are the four velocity at $x'$ and $x''$
respectively.
Further I define Synge's world function $\sigma(x, \bar x)$ which is
numerically equal to half the squared
geodesic distance between two points $x$ and $\bar x$. Using its gradient
$\sigma_\alpha = \nabla_\alpha \sigma(x, \bar x)$, I
define $r = u^{\alpha'} \sigma_{\alpha'}(x, x')$ and $r_{\text{adv}} =
-u^{\alpha''} \sigma_{\alpha''}(x, x'')$, the affine parameter distances of $x$ 
away from the world line along its future/past light cone. The potentials $U$
and $V$ appearing in Eq.~\eqref{eqn:singular-vector-gradient} are the direct and tail
parts of the retarded Green function $G_{\alpha
\bar \beta}(x, \bar x)$ associated with the wave operator.

From the definition of $r$, $\radv$, $u$, and $v$ it follows that (see Section~3.3.3
of~\cite{poisson:04})
\begin{gather}
    \nabla_\alpha u = -\sigma_\alpha(x,x')/r \label{eqn:subst-begin}
    \text{,}
    \\
    \nabla_\alpha v =  \sigma_\alpha(x,x'')/\radv
    \text{,}
    \\
    \nabla_\alpha r = -\sigma_{\alpha'\beta'}u^{\alpha'}u^{\beta'}
      \nabla_\alpha u + \sigma_{\alpha'\alpha} u^{\alpha'}
    \text{,}
    \\
    \nabla_\alpha \radv = -\sigma_{\alpha''\beta''}u^{\alpha''}u^{\beta''}
      \nabla_\alpha u - \sigma_{\alpha''\alpha} u^{\alpha''}
    \text{,}
\end{gather}
which are valid for geodesic motion. 

The potentials $U_{\alpha\beta'}$, $U_{\alpha\beta''}$ are determined by
Eq.~(322) of~\cite{poisson:04}
\begin{gather}
    {U_\alpha}^{\beta'}  = \pp{\beta' }{\alpha} \Delta^{1/2}(x,x')
    \text{,}
    \\
    {U_\alpha}^{\beta''} = \pp{\beta''}{\alpha} \Delta^{1/2}(x,x'')
    \text{,}\label{eqn:U-potential}
\end{gather}
where $\pp{\bar{\mu}}{\nu}$ is the parallel propagator from $x^\nu$ to $\bar
x^\mu$ and
$\Delta \define \det\bm(-\pp{\alpha'}{\alpha} \sigma^\alpha_{;\beta'} \bm)$
is the van~Vleck determinant.
Of the potentials $V_{\alpha\beta'}$ and $V_{\alpha\beta''}$ appearing in
Eq.~\eqref{eqn:singular-vector-gradient} I only need to know the
scaling behaviour following from Eq.~(320) of~\cite{poisson:04}:
\begin{gather}
    V_{\alpha\beta'} = \orderof{\varepsilon^2}
    \text{,}
    \\
    V_{\alpha\beta''} = \orderof{\varepsilon^2}
    \text{,}
    \\
    \nabla_\beta V_{\alpha\mu} = \orderof{\varepsilon}
    \text{.}
\end{gather}
These expressions are valid in vacuum spacetimes where the Ricci tensor vanishes.

Again mirroring the calculation in~\paperI, I introduce the arbitrary
point $\bar x \define z(\bar \tau)$ on the world line and expand the
quantities in Eq.~\eqref{eqn:singular-vector-gradient} in terms of
a Taylor expansion around $\bar x$.
I introduce the convenient quantities 
\begin{gather}
    \bar r \define \sigma_{\bar \alpha}(x, \bar x) u^{\bar \alpha}
    \text{,}
    \\
    s^2 \define (g^{\bar\alpha\bar\beta}+u^{\bar\alpha}u^{\bar\beta})
      \sigma_{\bar\alpha}(x,\bar x)\sigma_{\bar\beta}(x,\bar x)
    \text{,}
\end{gather}
together with the time differences
\begin{equation}
    \Delta_+ \define v - \bar\tau\text{,} \qquad \Delta_- \define u - \bar\tau
\end{equation}
from the advanced (retarded) point to the reference point $\bar x$.

I also use the expansion of the derivatives of the
parallel propagator around the point $\bar x$
\begin{gather}
    \pp{\bar\alpha}{\beta;\bar\gamma} = -\pp{\bar \beta}{\beta}
    \left( 
    \frac12 \Riemann{\bar\alpha}{\bar\beta\bar\gamma\bar\delta}\sigma^{\bar\delta}
    -\frac16 \Riemann{\bar\alpha}{\bar\beta\bar\gamma\bar\delta;\bar\varepsilon}\sigma^{\bar\delta}\sigma^{\bar\varepsilon}
    \right) + \orderof{\varepsilon^3}
    \text{,}
    \\
    \begin{split}
    \pp{\bar\alpha}{\beta;\gamma} &= - \pp{\bar\beta}{\beta} \pp{\bar\gamma}{\gamma} 
    \left( 
    \frac12 \Riemann{\bar\alpha}{\bar\beta\bar\gamma\bar\delta}\sigma^{\bar\delta}
    -\frac13 \Riemann{\bar\alpha}{\bar\beta\bar\gamma\bar\delta;\bar\varepsilon}\sigma^{\bar\delta}\sigma^{\bar\varepsilon}
    \right) 
     \\ & \quad 
    +\orderof{\varepsilon^3}
    \text{,} 
    \end{split}
\end{gather}
as well as an expansion for the second derivative of Synge's world function
\begin{multline}
    \sigma_{\bar\alpha\bar\beta} = g_{\bar\alpha\bar\beta} 
      - \frac13 R_{\bar\alpha\bar\gamma\bar\beta\bar\delta}\sigma^{\bar\gamma}\sigma^{\bar\delta}
      \\
      + \frac{1}{12} R_{\bar\alpha\bar\gamma\bar\beta\bar\delta;\bar\varepsilon}\sigma^{\bar\gamma}\sigma^{\bar\delta}\sigma^{\bar\varepsilon}
      +\orderof{\varepsilon^4}
      \text{,}
\end{multline}
and the van~Vleck determinant
\begin{equation}
    \Delta^{1/2} = 1 + \orderof{\epsilon^4}
    \text{,}
\end{equation}
which I calculate using the methods described in Sec.~(2.4.2)
of~\cite{poisson:04}.
    
I make use of the fact that the bi\nobreakdash-tensors 
\begin{gather}
    U_\alpha(\tau) \define U_{\alpha\mu} u^{\mu} 
    \text{,}
    \\
    U_{\alpha\beta}(\tau) \define U_{\alpha\mu;\beta} u^{\mu}
    \text{, and}
    \\
    \dot U_\alpha(\tau) \define U_{\alpha\mu;\nu} u^{\mu} u^{\nu}
    \label{eqn:U-alpha-expansion} 
\end{gather}
appearing in
Eq.~\eqref{eqn:singular-vector-gradient} do not bear a free index
on the world line, making them scalars on the world line. With $w$ being
either $u$ or $v$ and $\Delta \define w-\bar\tau = \Delta_\mp$ I expand these as 
\begin{gather}
    U_\alpha(w) = U_\alpha + \dot U_\alpha \Delta + \frac12 \ddot U_\alpha
      \Delta^2 + \frac16 U_\alpha^{(3)} \Delta^3 + \orderof{\varepsilon^4}
    \text{,}
    \\
    U_{\alpha\beta}(w) = U_{\alpha\beta} + \dot U_{\alpha\beta} \Delta + 
      \frac12 \ddot U_{\alpha\beta} \Delta^2 + \orderof{\varepsilon^3}
    \text{, and}
    \\
    \dot U_\alpha(w) = \dot U_\alpha + \ddot U_\alpha \Delta + 
      \frac12 U_\alpha^{(3)} \Delta^2 + \orderof{\varepsilon^3}
    \text{,}
\end{gather}
where it is understood that the coefficient functions are evaluated at $\tau = 
\bar \tau$.

Repeatedly taking derivatives of Eq.~\eqref{eqn:U-potential} 
and contracting with $u^{\bar \mu}$ I find for
the first set of coefficients
\begin{gather}
    U_\alpha = \pp{\bar\alpha}{\alpha} u_{\bar\alpha} + \orderof{\varepsilon^4}
    \text{,}
    \\
    \dot U_\alpha = \pp{\bar \alpha}{\alpha} \left( 
      \frac12 R_{\bar\alpha u u\sigma} 
     -\frac16 R_{\bar\alpha u u \sigma | \sigma} \right)
     +\orderof{\varepsilon^3}
    \text{,}
    \\
    \ddot U_\alpha = \frac13 \pp{\bar \alpha}{\alpha} R_{\bar \alpha u u \sigma | u}
     +\orderof{\varepsilon^2}
    \text{,}
    \\
    U^{(3)}_\alpha = 0 + \orderof{\varepsilon}
    \text{,}
\end{gather}
where I have introduced the notation 
$R_{\bar\alpha u u\sigma} \define R_{\bar\alpha \bar\beta \bar\gamma
\bar\delta} u^{\bar\beta} u^{\bar\gamma} \sigma^{\bar\delta}$ and 
$R_{\bar\alpha u u \sigma | \sigma} \define R_{\bar\alpha \bar\beta \bar\gamma
\bar\delta ; \bar \varepsilon} u^{\bar\beta} u^{\bar\gamma}
\sigma^{\bar\delta} \sigma^{\bar\varepsilon}$; I will use this notation and
its natural extension to higher derivatives and different combinations of
$u^{\bar\mu}$ and $\sigma^{\bar\mu}$ frequently below. 

Similarly I find for the second set 
\begin{gather}
    U_{\alpha\beta} = \pp{\bar\alpha}{\alpha} \pp{\bar\beta}{\beta} \left( 
    \frac12 R_{\bar\alpha u \bar\beta \sigma} 
    -\frac13 R_{\bar\alpha u \bar\beta \sigma | \sigma}
    \right) +\orderof{\varepsilon^3}
    \text{,}
    \\
    \begin{split}
    \dot U_{\alpha\beta} &= \pp{\bar\alpha}{\alpha} \pp{\bar\beta}{\beta} \left(
    \frac12 R_{\bar\alpha u \bar\beta u} 
    +\frac16 R_{\bar\alpha u \bar\beta \sigma | u}
    -\frac13 R_{\bar\alpha u \bar\beta u | \sigma}
    \right) 
    \\&\quad
    +\orderof{\varepsilon^2}
    \text{,}
    \end{split}
    \\
    \ddot U_{\alpha\beta} = \frac13 \pp{\bar\alpha}{\alpha} \pp{\bar\beta}{\beta} 
    R_{\bar\alpha u \bar\beta u | u}
    +\orderof{\varepsilon}\text{.}
\end{gather}
Note that the third set does not involve new coefficients, but only those
already calculated for $U_\alpha$.

Finally I copy expressions for $\Delta_\pm$, $r$, $\radv$, $u$, $v$ and their
gradients from~\paperI
\begin{widetext}
\begin{gather}
    \Delta_\pm = (\bar r \pm s) \mp \frac{(\bar r \pm s)^2}{6s} 
    R_{u \sigma u \sigma} \mp \frac{(\bar r \pm s)^2}{24s}\left[ 
    (\bar r \pm s) R_{u \sigma u \sigma | u} - R_{u \sigma u \sigma | \sigma}
    \right] + \orderof{\varepsilon^5}
    \text{,}
    \\
    r = s - \frac{\bar r^2 - s^2}{6s}R_{u \sigma u \sigma} -
    \frac{\bar r - s}{24 s}\left[ (\bar r - s)(\bar r + 2s) R_{u \sigma u
    \sigma | u} - (\bar r + s) R_{u \sigma u \sigma | \sigma} \right] +
    \orderof{\varepsilon^5}
    \text{,}
    \\
    \radv = s - \frac{\bar r^2 - s^2}{6s}R_{u \sigma u \sigma} -
    \frac{\bar r + s}{24 s}\left[ (\bar r + s)(\bar r - 2s) R_{u \sigma u
    \sigma | u} - (\bar r - s) R_{u \sigma u \sigma | \sigma} \right] +
    \orderof{\varepsilon^5}
    \text{,}
    \\
    \begin{split}
    \nabla_\alpha u &= \frac1s \pp{\bar \alpha}{\alpha} \biggl\{ 
      \left[ 
        \sigma_{\bar \alpha} + (\bar r - s) u_{\bar \alpha} 
      \right] 
      \\&\quad
      +\left[ 
        \frac16 (\bar r - s) R_{\bar\alpha \sigma u \sigma} - 
	\frac13 (\bar r - s)^2 R_{\bar\alpha u \sigma u} + 
	\frac{\bar r^2 - s^2}{6s^2} R_{u \sigma u \sigma} \sigma_{\bar\alpha} + 
	\frac{(\bar r - s)^2(\bar r + 2s)}{6s^2} R_{\sigma u \sigma u | u} u_{\bar\alpha} 
      \right] 
      \\&\quad
      +\biggl[ 
        -\frac{1}{12}(\bar r - s) R_{\bar\alpha u \sigma u | \sigma} +
	\frac18 (\bar r - s)^2 R_{\bar\alpha u \sigma u | \sigma} +
        \frac{1}{24} (\bar r - s)^2 R_{\bar \alpha \sigma u \sigma | u}
        -\frac{1}{12} (\bar r - s)^3 R_{\bar\alpha u \sigma u | u} 
	\\&\quad
	+ \frac{\bar r - s}{24s^2} 
	\left( 
	  (\bar r - s)(\bar r +2s) R_{u \sigma u \sigma | u} -
          (\bar r + s) R_{u \sigma u \sigma | \sigma} 
	\right) \sigma_{\bar \alpha} 
	\\&\quad
	+\frac{(\bar r - s)^2}{24s^2} 
	\left( 
	  (\bar r - s)(\bar r +3s) R_{u \sigma u \sigma | \sigma} -
	  (\bar r + 2s) R_{u \sigma u \sigma | \sigma}
	\right) u_{\bar \alpha}
      \biggr] +
      \orderof{\varepsilon^5} 
    \biggr\}
    \text{,}
    \end{split}
    \\
    \begin{split}
    \nabla_\alpha v &= -\frac1s \pp{\bar \alpha}{\alpha} \biggl\{ 
      \left[ 
        \sigma_{\bar \alpha} + (\bar r + s) u_{\bar \alpha} 
      \right] 
      \\&\quad
      +\left[ 
        \frac16 (\bar r + s) R_{\bar\alpha \sigma u \sigma} - 
	\frac13 (\bar r + s)^2 R_{\bar\alpha u \sigma u} + 
	\frac{\bar r^2 - s^2}{6s^2} R_{u \sigma u \sigma} \sigma_{\bar\alpha} + 
	\frac{(\bar r + s)^2(\bar r - 2s)}{6s^2} R_{\sigma u \sigma u | u} u_{\bar\alpha} 
      \right] 
      \\&\quad
      +\biggl[ 
        -\frac{1}{12}(\bar r + s) R_{\bar\alpha u \sigma u | \sigma} +
	\frac18 (\bar r + s)^2 R_{\bar\alpha u \sigma u | \sigma} +
        \frac{1}{24} (\bar r + s)^2 R_{\bar \alpha \sigma u \sigma | u}
        -\frac{1}{12} (\bar r + s)^3 R_{\bar\alpha u \sigma u | u} 
	\\&\quad
	+ \frac{\bar r + s}{24s^2} 
	\left( 
	  (\bar r + s)(\bar r -2s) R_{u \sigma u \sigma | u} -
          (\bar r - s) R_{u \sigma u \sigma | \sigma} 
	\right) \sigma_{\bar \alpha} 
	\\&\quad
	+\frac{(\bar r + s)^2}{24s^2} 
	\left( 
	  (\bar r + s)(\bar r -3s) R_{u \sigma u \sigma | \sigma} -
	  (\bar r - 2s) R_{u \sigma u \sigma | \sigma}
	\right) u_{\bar \alpha}
      \biggr] +
      \orderof{\varepsilon^5} 
    \biggr\}
    \text{,}
    \end{split}
    \\
    \begin{split}
    % copied from Eric's paper
    \nabla_\alpha r &= -\frac1s \pp{\bar \alpha}{\alpha} \left\{   
      \left[ 
        \sigma_{\bar\alpha} + \bar r u_{\bar\alpha} 
      \right] 
      + \left[ 
        \frac16 \bar r R_{\bar\alpha \sigma u \sigma} 
        -\frac13 (\bar r^2-s^2) R_{\bar\alpha u \sigma u} 
        +\frac{\bar r ^2+s^2}{6s^2} R_{u\sigma u\sigma} \sigma_{\bar\alpha} 
        +\frac{\bar r (\bar r ^2-s^2)}{6s^2} R_{u\sigma u\sigma} u_{\bar\alpha}
      \right]  
      \right.\\&\quad\left.
      +\left[ 
        -\frac{1}{12} \bar r R_{\bar\alpha \sigma u \sigma|\sigma}  
        +\frac18 (\bar r^2-s^2) R_{\bar\alpha u \sigma u |\sigma}  
	+\frac{1}{24} (\bar r^2-s^2) R_{\bar\alpha \sigma u \sigma | u}  
        \right.\right.\\&\quad\left.\left.
	-\frac{1}{12} (\bar r-s)^2(\bar r+2s) R_{\bar\alpha u \sigma u | u}   
	+\frac{1}{24s^2} \left( 
          (\bar r-s)(\bar r^2+\bar r s+4s^2) R_{u \sigma u \sigma | u}  
          -(\bar r^2+s^2) R_{u \sigma u\sigma | \sigma}
	\right) \sigma_{\bar\alpha}  
        \right.\right.\\&\quad\left.\left.
	+\frac{\bar r-s}{24s^2} \left(
          (\bar r-s)(\bar r^2+2\bar r s+3s^2) R_{u \sigma u\sigma | u}  
          -\bar r (\bar r + s) R_{u\sigma u\sigma | \sigma}
	\right)  u_{\bar\alpha} 
      \right]  
      + \orderof{\epsilon^5}
    \right\}
    \text{,}
    \end{split}
    \\
    \begin{split}
    \nabla_\alpha \radv &= -\frac1s \pp{\bar \alpha}{\alpha} \left\{   
      \left[ 
        \sigma_{\bar\alpha} + \bar r u_{\bar\alpha} 
      \right] 
      + \left[ 
        \frac16 \bar r R_{\bar\alpha \sigma u \sigma} 
        -\frac13 (\bar r^2-s^2) R_{\bar\alpha u \sigma u} 
        +\frac{\bar r ^2+s^2}{6s^2} R_{u\sigma u\sigma} \sigma_{\bar\alpha} 
        +\frac{\bar r (\bar r ^2-s^2)}{6s^2} R_{u\sigma u\sigma} u_{\bar\alpha}
      \right]  
      \right.\\&\quad\left.
      +\left[ 
        -\frac{1}{12} \bar r R_{\bar\alpha \sigma u \sigma|\sigma}  
        +\frac18 (\bar r^2-s^2) R_{\bar\alpha u \sigma u |\sigma}  
	+\frac{1}{24} (\bar r^2-s^2) R_{\bar\alpha \sigma u \sigma | u}  
        \right.\right.\\&\quad\left.\left.
	-\frac{1}{12} (\bar r+s)^2(\bar r-2s) R_{\bar\alpha u \sigma u | u}   
	+\frac{1}{24s^2} \left( 
          (\bar r+s)(\bar r^2-\bar r s+4s^2) R_{u \sigma u \sigma | u}  
          -(\bar r^2+s^2) R_{u \sigma u\sigma | \sigma}
	\right) \sigma_{\bar\alpha}  
        \right.\right.\\&\quad\left.\left.
	+\frac{\bar r+s}{24s^2} \left(
          (\bar r+s)(\bar r^2-2\bar r s+3s^2) R_{u \sigma u\sigma | u}  
          -\bar r (\bar r - s) R_{u\sigma u\sigma | \sigma}
	\right)  u_{\bar\alpha} 
      \right]  
      + \orderof{\epsilon^5}
      \right\}\label{eqn:subst-end}
    \text{.}
    \end{split}
\end{gather}
\end{widetext}

After substituting Eqs.~\eqref{eqn:subst-begin} --
\eqref{eqn:subst-end} into
Eq.~\eqref{eqn:singular-vector-gradient} (all of them) and sorting
out the orders I find the final expression for the covariant expansion of
$A^S_{\alpha;\beta}$
\begin{widetext}
\begin{multline}
    A^S_{\alpha;\beta} =
    q \pp{\bar \alpha}{\alpha} \pp{\bar \beta}{\beta}
    \left\{
    \left[ 
    \frac{1}{s^3} u_{\bar\alpha} \sigma_{\bar\beta}
    +\frac{\bar r}{s^3} u_{\bar\alpha} u_{\bar\beta}
    \right]
    +\left[
    \frac{\bar r}{6s^3} u_{\bar\alpha} R_{\bar\beta \sigma u \sigma}
    +\left( 
      \frac{\bar r}{2s^3} \sigma_{\bar\beta}
      +\frac{\bar r^2-s^2}{2s^3} u_{\bar\beta}
    \right) R_{\bar\alpha u u \sigma}
    \right.\right.\\\mbox{}\left.\left.
    +\frac{\bar r^2-s^2}{3s^2} u_{\bar\alpha} R_{\bar\beta u u \sigma}
    +\frac{1}{2s} R_{\bar\alpha u \bar\beta \sigma}
    +\frac{3\bar r^2-s^2}{6s^5} R_{u \sigma u \sigma} u_{\bar\alpha} \sigma_{\bar\beta}
    +\frac{\bar r(\bar r^2-s^2)}{2s^5} R_{u \sigma u \sigma} u_{\bar\alpha} u_{\bar\beta}
    +\frac{\bar r}{2s} R_{\bar\alpha u \bar\beta u}
    \right]
    \right.\\\mbox{}\left.
    +\left[ 
    -\frac{\bar r}{12 s^3} u_{\bar\alpha} R_{\bar \beta \sigma u \sigma | \sigma}
    -\frac{\bar r^2 - s^2}{24s^3} u_{\bar\alpha} R_{\bar\beta \sigma \sigma u | u}
    -\left( 
      \frac{\bar r}{6s^3} \sigma_{\bar\beta}
      +\frac{\bar r^2-s^2}{6s^3} u_{\bar\beta}
    \right)R_{\bar\alpha u u \sigma | \sigma}
    -\frac{\bar r^2-s^2}{8s^3} u_{\bar\alpha} R_{\bar\beta u u \sigma | \sigma}
    \right.\right.\\\mbox{}\left.\left.
    -\frac{1}{3s^2} R_{\bar\alpha u \bar\beta \sigma | \sigma}
    +\left( 
      \frac{\bar r^2-s^2}{6s^3} \sigma_{\bar\beta} 
      +\frac{\bar r (\bar r^2 - 3s^2)}{6s^3} u_{\bar\beta}
    \right) R_{\bar\alpha u u \sigma | u}
    +\frac{\bar r (\bar r^2 -3s^2)}{12s^3}u_{\bar\alpha} R_{\bar\beta u u \sigma | u}
    -\frac{\bar r}{3s} R_{\bar\alpha u \bar\beta u | \sigma}
    \right.\right.\\\mbox{}\left.\left.
    +\frac{\bar r}{6s} R_{\bar a u \bar b \sigma | u}
    +\frac{\bar r^2+s^2}{6s^3} R_{\bar a u \bar b u | u}
    +\left( 
      -\frac{3\bar r^2-s^2}{24s^5} R_{u \sigma u \sigma | \sigma} 
      +\frac{\bar r (\bar r^2-s^2)}{8s^5} R_{u \sigma u \sigma | u}
    \right) u_{\bar\alpha} \sigma_{\bar\beta}
    \right.\right.\\\mbox{}\left.\left.
    +\left( 
      -\frac{\bar r (\bar r^2-s^2)}{8s^5} R_{u \sigma u \sigma | \sigma}
      +\frac{(\bar r^2 -s^2)^2}{8s^5} R_{u \sigma u \sigma | u}
      \right) u_{\bar\alpha} u_{\bar\beta}
      \right]\right\}
    +\orderof{\varepsilon^2}
    \text{,}\label{eqn:covariant-expansion-singular-vector-gradient}
\end{multline}
\end{widetext}
where terms in square brackets are of the same power in $\varepsilon$.

I copy the
results for the coordinate expansion of $\sigma_{\bar\alpha}(x, \bar x)$ and
$\pp{\bar\alpha}{\beta}(x, \bar x)$ from Eqs.~(3.16) -- (3.19) and
Eqs.~(3.30) -- (3.33) of \paperI. I use
% copied from Eric's paper
\begin{gather} 
    \begin{split}
    -\sigma_{\bar\alpha}(x, \bar x) &= g_{\alpha\beta}w^\beta +
    A_{\alpha\beta\gamma} w^\beta w^\gamma +
    A_{\alpha\beta\gamma\delta} w^\beta w^\gamma w^\delta 
    \\&\mbox{}
    + A_{\alpha\beta\gamma\delta\varepsilon} w^\beta w^\gamma w^\delta w^\varepsilon
    + \orderof{\varepsilon^5}
    \text{,}
    \end{split} 
    \\
    \A{\alpha}{\beta\gamma} \define \frac{1}{2} \Chr{\alpha}{\beta\gamma}
    \text{,}
    \\
    \A{\alpha}{\beta\gamma\delta} \define \frac{1}{6} 
    \left( 
      \Chr{\alpha}{\beta\gamma,\delta} 
      + \Chr{\alpha}{\beta\mu} \Chr{\mu}{\gamma\delta} 
    \right)\text{,}
    \\
    \begin{split}
    \A{\alpha}{\beta\gamma\delta\epsilon} &\define \frac{1}{24} 
    \left( 
      \Chr{\alpha}{\beta\gamma,\delta\epsilon} 
      + \Chr{\alpha}{\beta\gamma,\mu} \Chr{\mu}{\delta\epsilon} 
      \right.\\&\mbox{}\left.
      + 2 \Chr{\alpha}{\beta\mu} \Chr{\mu}{\gamma\delta,\epsilon} 
      + \Chr{\alpha}{\mu\nu} \Chr{\mu}{\beta\gamma} 
      \Chr{\nu}{\delta\epsilon} 
    \right)\text{,}
    \end{split}
\end{gather}
as well as
\begin{gather}
    \begin{split}
    \pp{\bar \mu}{\alpha}(x,\bar x) &= {\delta^\mu}_{\alpha} 
    + \B{\mu}{\alpha\beta} w^\beta  
    + \B{\mu}{\alpha\beta\gamma} w^\beta w^\gamma 
    \\&\mbox{}
    + \B{\mu}{\alpha\beta\gamma\delta} w^\beta w^\gamma w^\delta  
    + \orderof{\varepsilon^4}\text{,}
    \text{,}
    \end{split} 
    \\
    \B{\mu}{\alpha\beta} \define \Chr{\mu}{\alpha\beta}\text{,}
    \\
    \B{\mu}{\alpha\beta\gamma} \define \frac12 \left( 
      \Chr{\mu}{\alpha\beta,\gamma} 
      +\Chr{\mu}{\beta\nu} \Chr{\nu}{\alpha\gamma} \right)
    \text{,} 
    \\
    \begin{split}
    \B{\mu}{\alpha\beta\gamma\delta} &\define \frac{1}{12} \left( 
      2\Chr{\mu}{\alpha\beta,\gamma\delta} 
      +2\Chr{\nu}{\alpha\beta} \Chr{\mu}{\nu\gamma,\delta}
      \right.\\&\mbox{}\left.
      -\Chr{\nu}{\beta\gamma} \Chr{\mu}{\alpha\nu,\delta} 
      +4\Chr{\mu}{\beta\nu} \Chr{\nu}{\alpha\gamma,\delta}
      \right.\\&\mbox{}\left.
      +\Chr{\nu}{\beta\gamma} \Chr{\mu}{\alpha\delta,\nu} 
      -\Chr{\mu}{\beta\nu} \Chr{\nu}{\alpha\lambda} \Chr{\lambda}{\gamma\delta}
      \right.\\&\mbox{}\left.
      +\Chr{\mu}{\nu\lambda} \Chr{\nu}{\alpha\beta} \Chr{\lambda}{\gamma\delta} 
      +2\Chr{\mu}{\beta\nu} \Chr{\nu}{\gamma\lambda} \Chr{\lambda}{\alpha\delta} 
    \right)\text{,}
    \end{split}
\end{gather}
where $w^\alpha \define x^\alpha - x^{\bar\alpha}$ is the coordinate distance
between $x$ and $\bar x$. Together with
Eq.~\eqref{eqn:covariant-expansion-singular-vector-gradient} these equations
form an expansion in $w^\alpha$ of the singular part of the gradient of the
vector potential around a point $x$ near the world line of the particle. I
finally calculate the tetrad components of the singular Faraday tensor as
\begin{equation}
    F^S_{(\mu)(\nu)} = (A^S_{\beta,\alpha}-A^S_{\alpha,\beta}) 
      e^\alpha_{\ (\mu)} e^\beta_{\ (\nu)}
    \text{.}
\end{equation}
From this point on I proceed exactly as described in Section~V
of~\paperI{} using \programname{Maple} and \programname{GRTensorII} to
perform the calculations. I find, after an extremely tedious calculation,
\begin{widetext}
\begin{gather}
	A_{(0)(+)} = \sign(\Delta) \Bigl[ 
	\frac{i \dot r_0 J}{r_0 \frakf a^2}
	-\frac{1}{r_0^2} \Bigr] e^{i\varphi_0}
    \text{,}\\
	B_{(0)(+)} = \biggl\{
	\Bigl[ 
	  -\frac{i E (J^2 - r_0^2) \dot r_0}{a^3 \pi \frakf  J}
	  +\frac{E(2 - \frakf )}{\pi \frakf  a r_0}
	\Bigr] \mathcal{E}
	- \frac{i r_0^2 E \dot r_0}{a^3 J \frakf  \pi} \mathcal{K}
	\biggr\} e^{i\varphi_0}
    \text{,}\\
    \begin{split}
	D&_{(0)(+)} = \Bigg\{
	\biggl[
	\frac{i E r_0^2 (-14r_0^2 J^2 + J^4 + r_0^4) \dot r_0^3}{8\pi J \frakf  a^7}
	-\frac{(-r_0 \frakf  J^2 + 2r_0J^2 + 7r_0^3 \frakf  - 14r_0^3) E \dot r_0^2}{8a^5 \frakf  \pi}
	\\&\mbox{}
	+i \frac{\bigl(8MJ^8-14MJ^6r_0^2-3r_0^5J^4-80MJ^4r_0^4+4J^4r_0^5 \frakf -7r_0^7J^2-68Mr_0^6J^2+4r_0^9-26Mr_0^8-4r_0^9 \frakf \bigr)E\dot r_0}{8r_0^5a^5\frakf J\pi}
	\\&\mbox{}
	-\frac{(8Mr_0\frakf J^6-8r_0^3MJ^4+38J^4r_0^3\frakf -2r_0^6J^2-16Mr_0^5J^2+3J^2r_0^6\frakf+54J^2r_0^5\frakf+20r_0^7\frakf +5r_0^8\frakf -6r_0^8)E}{8r_0^7a^3\frakf \pi}
	\biggr] \mathcal{E}
	\\&\mbox{}
	+\biggl[
	\frac{i E r_0^4(7J^2-r_0^2)\dot r_0^3}{8\pi J\frakf a^7}
	-\frac{(2-\frakf )r_0^3E\dot r_0^2}{2a^5\frakf \pi}
	\\&\mbox{}
	+\frac{\bigl(4Mr_0\frakf J^4+20J^2r_0^3\frakf +8Mr_0^3J^2+14r_0^5M\frakf -2r_0^6+12Mr_0^5+r_0^6\frakf \bigr)E}{8r_0^5a^3\frakf \pi}
	\\&\mbox{}
	-\frac{i(2MJ^6-9Mr_0^2J^4-2J^2r_0^5\frakf -20Mr_0^4J^2-2r_0^7\frakf +2r_0^7-13Mr_0^6)E\dot r_0}{4r_0^3a^5\frakf J\pi} 
	\biggr] 
	\mathcal{K}
	\Bigg\} e^{i\varphi_0}
	\text{,}
    \end{split}
    \\
	A_{(+)(-)} = 
	\sign(\Delta)\frac{2iEJ}{a^2r_0\frakf } 
	e^{i\varphi_0}
    \text{,}\\
    \begin{split}
	B_{(+)(-)} &= -2i \Biggl\{
	\biggl[ 
	-\frac{(r_0^2-J^2)\dot r_0^2}{a^3\pi J \frakf }
	+\frac{-J^2r_0\frakf +2r_0J^2+2r_0^3-2r_0^3\frakf }{r_0^3aJ\pi}
	\biggr]\mathcal{E}
	+\biggl[ 
	\frac{r_0^2\dot r_0^2}{a^3\frakf J\pi}
	-\frac{2(1-\frakf )}{aJ\pi}
	\biggr]\mathcal{K}
	\Biggr\} e^{i\varphi_0}
	\text{,}
    \end{split}
    \\
    \begin{split}
	D_{(+)(-)} &= -2i \Bigg\{
	\biggl[
	-\frac{r_0^2(-14r_0^2J^2+J^4+r_0^4)\dot r_0^4}{8\frakf \pi Ja^7}
	-\Bigl(4M\frakf J^8-7Mr_0^2\frakf J^6+2J^4r_0^5\frakf +2J^4r_0^4M-J^4r_0^5
	  \\&\mbox{}
	  -43J^4r_0^4\frakf M-7J^2r_0^7\frakf -27J^2Mr_0^6\frakf -11Mr_0^8\frakf -r_0^9\frakf +r_0^9-2r_0^8M\Bigr)
	  r_0^{1/2} \dot r_0^2 \Big/ \Bigl(4r_0^5a^5J\frakf \pi\Bigr)
	\\&\mbox{}
	-\Bigl(8M\frakf J^8-8J^6Mr_0^2+30Mr_0^2\frakf J^6-2J^4r_0^5+10J^4r_0^4\frakf M-24J^4r_0^4M+3J^4r_0^5\frakf 
	  \\&\mbox{}
	  -28J^2Mr_0^6+J^2r_0^7\frakf -28J^2Mr_0^6\frakf -20Mr_0^8\frakf -12r_0^8M\Bigr)
	  \Big/ \Bigl(8r_0^7a^3J\pi\Bigr)
	\biggr]\mathcal{E}
	\\&\mbox{}
	+\biggl[
	-\frac{r_0^4(7J^2-r_0^2)\dot r_0^4}{8\frakf \pi Ja^7}
	  \\&\mbox{}
	+\frac{4M\frakf J^6-16J^4Mr_0^2+4r_0^2\frakf MJ^4-18J^2r_0^4\frakf M-28J^2r_0^4M-J^2\frakf r_0^5-12r_0^6M-20r_0^6\frakf M}{8r_0^5a^3J\pi}
	  \\&\mbox{}
	+ %\frac{
	  \Bigl(2M\frakf J^6-9r_0^2\frakf MJ^4+J^2r_0^5-2J^2r_0^4M-5J^2\frakf r_0^5-14J^2r_0^4\frakf M-2r_0^6M+r_0^7
	  \\&\mbox{}
	  -11r_0^6\frakf M-\frakf r_0^7\Bigr)\dot r_0^2 \Big/
	  %}{
	  \Bigl(
	  4r_0^{5/2}a^5J\frakf\pi\Bigr)
	  %}
	\biggr]\mathcal{K}
	\Bigg\}e^{i\varphi_0}
	\text{,}
    \end{split}
\end{gather}
where $\frakf = \sqrt{\frac{r_0-2M}{r_0}}$, $a^2 = r_0^2 + J^2$.
\end{widetext}
%
% copied from Eric's paper
Here, the rescaled elliptic integrals $\mathcal{E}$ and $\mathcal{K}$ are
defined by
\begin{equation} 
\mathcal{E} \define \frac{2}{\pi} \int_0^{\pi/2} (1-k\sin^2\psi)^{1/2}\, \d\psi  
= F\left(-{\frac{1}{2}},{\frac{1}{2}}; 1; k\right)
    \text{,}
\end{equation} 
and 
\begin{equation} 
\mathcal{K} \define \frac{2}{\pi} \int_0^{\pi/2} (1-k\sin^2\psi)^{-1/2}\, \d\psi     
= F\left({\frac{1}{2}},{\frac{1}{2}}; 1; k\right)\text{,} 
\end{equation} 
in which $F(a,b;c;x)$ are the hypergeometric functions and $k \define J^2/(r_0^2 + J^2)$. 

\section{Vector potential calculation\label{sec:vector-pot-calculation}}
In this section I describe a variant of the numerical calculation discussed
in the main part of the paper that uses the vector potential instead of the
Faraday tensor. To this end I decompose the vector
potential and the sources in terms of vectorial spherical harmonics
\begin{subequations}
\label{eqn:vector-pot-decomp}
\begin{align}
    A_a(t,r^*,\theta,\phi) &= {\textstyle \frac{1}{r}} A^{\ell m}_a(t, r^*) Y_{\ell  m}(\theta, \phi) 
    \text{,}
    \\
    j_a(t,r^*,\theta,\phi) &= j^{\ell m}_a(t, r^*) Y_{\ell m}(\theta, \phi)
      & \text{for $a = t,r^*$,}
    \\
    A_A(t,r^*,\theta,\phi) &= v_{\ell m}(t, r^*) Z_A^{\ell m}(\theta, \phi) 
    \nonumber\\&\quad
    + \tilde v_{\ell m}(t, r^*) X_A^{\ell m}(\theta, \phi)
    \text{,}
    \\
    j_A(t,r^*,\theta,\phi) &= j^{\text{even}}_{\ell m}(t, r^*) Z_A^{\ell m}(\theta, \phi) 
    \nonumber\\&\quad
    + j^{\text{odd}}_{\ell m}(t, r^*) X_A^{\ell m}(\theta, \phi)
      & \text{for $A = \theta,\phi$,}
\end{align}
\end{subequations}
and substitute this into the Maxwell equations for the vector potential in the
Lorenz gauge $g^{\alpha\beta} A_{\alpha;\beta} = 0$:
\begin{equation}
    g^{\mu\nu} A_{\alpha;\mu\nu} - {R^{\beta}}_{\alpha} A^\beta = - 4\pi
    j_\alpha
    \text{,}\label{eqn:covariant-wave-eqn}
\end{equation}
where $R_{\alpha\beta}$ is the spacetime's Ricci tensor, which vanishes in
Schwarzschild spacetime. 
Substituting Eq.~\eqref{eqn:vector-pot-decomp} into
Eq.~\eqref{eqn:covariant-wave-eqn} I arrive at two decoupled sets of
equations for the even ($A^{\ell m}_a$, $v_{\ell m}$) and odd ($\tilde v_{\ell
m}$) modes
\begin{widetext}
\begin{gather}
    -\diffz{A^{\ell m}_t}{t} + \diffz{A^{\ell m}_t}{r^*}
      + \frac{2M}{r^2} \left( \diff{A^{\ell m}_{r^*}}{t}
      - \diff{A^{\ell m}_t    }{r^*} \right)
      - V A^{\ell m}_t = -4 \pi r f j^{\ell m}_t
      \label{eqn:reduced-wave-eqn-begin}
    \text{,}
    \\
    \begin{split}
    -\diffz{A^{\ell m}_{r^*}}{t} + \diffz{A^{\ell m}_{r^*}}{r^*} 
      + \frac{2M}{r^2} \left( \diff{A^{\ell m}_t    }{t}
      - \diff{A^{\ell m}_{r^*}}{r^*} \right) 
      - \left( V + 2 \frac{f^2}{r^2} \right) A^{\ell m}_{r^*} 
      + f V v_{\ell m} 
      = -4 \pi r f j^{\ell m}_{r^*}
    \text{,}
    \end{split}
    \\
    -\diffz{v_{\ell m}}{t} + \diffz{v_{\ell m}}{r^*}
      - V v_{\ell m} 
      + 2 \frac{f}{r^2} A^{\ell m}_{r^*} 
      = -4 \pi f j^{\text{even}}_{\ell m}
    \text{,}
    \\
    -\diffz{\tilde v_{\ell m}}{t} + \diffz{\tilde v_{\ell m}}{r^*}
      - V \tilde v_{\ell m} 
      = -4 \pi f j^{\text{odd}}_{\ell m}
    \text{,}\label{eqn:reduced-wave-eqn-end}
\end{gather}
\end{widetext}
where $V$ and $j^{\ell m}_\alpha$ is defined as in
Eqs.~\eqref{eqn:red-wave-psi} and~\eqref{eqn:red-sources} in the
main text.

\subsection{Numerical method}
I discretize the set of reduced equations
Eqs.~\eqref{eqn:reduced-wave-eqn-begin} -- \eqref{eqn:reduced-wave-eqn-end}
using Lousto's method as described in section~\ref{sec:numerical-method} of
the main text. Since the source terms on the right hand side are less singular
for the vector potential than they are for the Faraday tensor, I do not
have to distinguish between sourced and vacuum cells in the integral over the
potential terms. 

Terms containing first derivatives $\diff{\psi}{t}$, $\diff{\psi}{r^*}$, where 
now and in the remainder of the appendix $\psi$ stands for any of $A^{\ell
m}_t$, $A^{\ell m}_{r^*}$, $v^{\ell m}$ or $\tilde v^{\ell m}$,
were not treated in~\cite{lousto:97}, but, for generic vacuum
cells, can be handled in a straightforward manner
\begin{gather}
    \iint_{\text{\rlap{cell}}} \, \d u \, \d v \,  V(r) \diff{\psi}{t} = 
    2 h (\psi_3 - \psi_2) V_0 + \orderof{h^4}
    \label{eqn:dt-integral}
    \text{,}
    \\
    \iint_{\text{\rlap{cell}}} \, \d u \, \d v \,  V(r) \diff{\psi}{r^*} = 
    2 h (\psi_4 - \psi_1) V_0 + \orderof{h^4}
    \label{eqn:drstar-integral}\text{.}
\end{gather}
This fails for cells traversed by the particle, since the field is only
continuous across the world line but not differentiable. For these cells I
take recourse to Lousto's original algorithm, which has to deal with a similar
issue, and use
\begin{gather}
    \iint_{\text{\rlap{cell}}} \, \d u \, \d v \, V(r) \diff{\psi}{t} = 
    V_0 \, \sum_i A_i \partial_t \psi_i + \orderof{h^3}
    \label{eqn:sourced-dt-integral}
    \text{,}
    \\
    \iint_{\text{\rlap{cell}}} \, \d u \, \d v \, V(r) \diff{\psi}{r^*} = 
    V_0 \, \sum_i A_i \partial_{r^*} \psi_i
    + \orderof{h^3}
    \label{eqn:sourced-drstar-integral}\text{,}
\end{gather}
where $A_1$,\ldots,$A_4$ are the subareas indicated in
Fig.~\ref{fig:simpson-points} and $\partial_t \psi_1$, \ldots, $\partial_t
\psi_4$, $\partial_{r^*} \psi_1$, \ldots, $\partial_{r^*} \psi_4$
are zeroth order accurate
approximations to the derivatives in the subareas. I calculate these using
grid points outside of the cell on the same side of the world line as the
corresponding subarea, e.g.
\begin{equation}
\partial_{r^*} \psi_1 = \frac{\psi(t, r^*-h) -
\psi(t, r^* - 3 h)}{2h} + \orderof{h}\text{.} 
\end{equation}

\subsection{Gauge condition\label{sec:gauge-condition}}
In contrast to the scalar field, the electromagnetic vector potential has to
satisfy a gauge condition
\begin{equation}
    Z \define g^{\alpha \beta} A_{\alpha;\beta} = 0 \text{.}
\end{equation}
Analytically the gauge condition is preserved by the evolution equations, so
that it is sufficient to impose it on the initial data. Numerically, however,
small violations of the gauge condition due to the numerical approximation can
be amplified exponentially and come to dominate the numerical data. To handle
this situation I introduce a gauge damping scheme as described
in~\cite{gundlach:05, barack:05}. That is I add a term of the form 
\begin{multline}
    \frac{4M}{r^2} Z = \frac{4M}{r^2} \left(
    -\frac{1}{r-2M} \diff{A_t}{t} + \frac{1}{r-2M} \diff{A_{r^*}}{r^*} 
    \right.\\\mbox{}\left.
    + \frac{1}{r^2} A_{r^*} - \frac{\ell(\ell+1)}{r^2} v
    \right)
\end{multline}
to the $t$ components of the evolution equations
Eqs.~\eqref{eqn:covariant-wave-eqn}, which dampens out violations of the
gauge condition. This choice proved to be numerically stable for the
radiative ($\ell > 0$) modes but unstable for the monopole ($\ell = 0$) mode. 

\subsection{Monopole mode\label{sec:monopole-mode-vector}}
The monopole moment of an electromagnetic field is non-radiative. This
makes its behaviour sufficiently different from that of the radiative ($\ell >
0$) modes that the approach outlined earlier fails for $\ell = 0$. 
In this case Eq.~\eqref{eqn:covariant-wave-eqn} reduces to a set of
coupled equations for $A_a^{0,0}$ only. 
Rather than solving the system of equations directly for $A_t^{0,0}$ and
$A_{r^*}^{0,0}$ I use the analytical result for the $F_{tr}$ component of
the Faraday tensor derived in section~\ref{sec:monopole-mode} in the main part
of the paper. This proves to be sufficient to reconstruct the combination
$A^{0,0}_{r,t}-A^{0,0}_{t,r}$ appearing in Eq.~\eqref{eqn:translation-table}.

\subsection{Initial values and boundary conditions}
I handle the problem of initial data and boundary conditions the same way as
in the main text, that is I arbitrarily choose the fields to vanish on the
characteristic slices $u = u_0$ and $v = v_0$
\begin{equation}
    A_\alpha(u = u_0) = A_\alpha(v=v_0) = 0 \text{,}
\end{equation}
thereby adding a certain amount of spurious waves to the solution which show
up as an initial burst.
Gauge violations in this initial data are damped out along with those arising
during the evolution. 

I implement ingoing wave boundary conditions near the event horizon and
choose a numerical domain that covers the full domain of dependence of the 
initial data near the outer boundary.

\subsection{Extraction of the field data at the particle}
In order to extract the value of the fields and their first derivatives at the
position of the particle, I use a variant of the extraction scheme described
in~\paperII{}. I introduce a piecewise polynomial
\begin{equation}
    p(x)  = 
      \begin{cases}
	  c_0  + c_1 x  + \frac{c_3}{2} x^2 & \text{if $x < 0$}\\
	  c'_0 + c'_1 x + \frac{c'_3}{2} x^2 & \text{if $x > 0$}
      \end{cases}
\end{equation}
in $x \define r^* - r_0^*$ on the current slice. Its coefficients to the 
left and
right of the world line are linked by jump conditions 
$c_n = c'_n + \jump{\partial^n_{r^*} \psi}$ listed in
Appendix~\ref{sec:vector-jump-conditions}. Fitting this polynomial to the three grid
points closest to the particle, I extract approximations for $\psi(t_0,
r^*_0)$ and $\diff{\psi(t_0, r^*_0)}{r^*}$ which are just the coefficients
$c_0$, $c_1$ respectively. Once I have obtained these, I proceed as in
section~\ref{sec:field-extraction} of the main part of the paper
following~\cite{sago:07} to obtain values for $\diff{\psi(t_0, r^*_0)}{t}$.

\subsection{Results}
\begin{figure}
    \includegraphics{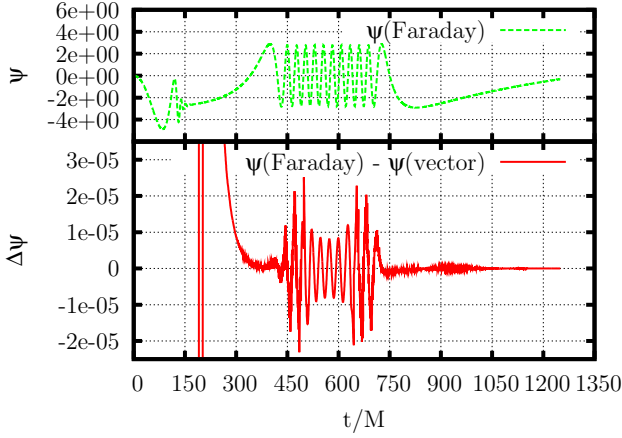}
    \caption{Differences between $F_{tr}^{\ell m}$ calculated using the vector
    potential and calculated using the Faraday tensor method for $\ell = 2$,
    $m = 2$ mode of field for the zoom-whirl orbit shown in
    Fig.~\protect\ref{fig:zoom-whirl-orbit-trajectory}. Displayed are the difference
    and the actual field. The stepsizes were $h = 1.041\bar6\times10^{-2}\,M$ and $h =
    1/512\,M$ for the vector potential calculation and the Faraday tensor
    calculation respectively.}\label{fig:vector-faraday-comparison}
\end{figure}

\begin{figure}
    \includegraphics{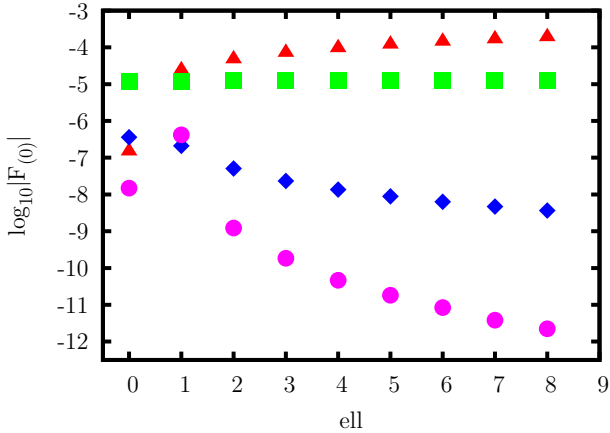}
    \caption{Multipole coefficients of $\frac{M^2}{q}\real F^R_{(0)}$ for a
    particle on a zoom-whirl orbit ($p = 7.8001$, $e = 0.9$), calculated using
    a stepsize of $h = 0.125M$ for the $\ell = 1$ modes and increasing the
    resolution linearly with $\ell$ for $\ell > 1$.
    The
    coefficients are extracted at $t = 1100\,M$ when the particle is deep
    within the zoom phase. Red triangles are used for the unregularized
    multipole coefficients $F_{(0),\ell}$, squares, diamonds and disks are
    used for the partly regularized coefficients after the removal of the
    $A_{(0)}$, $B_{(0)}$ and $D_{(0)}$ terms respectively.\label{fig:vector-multipole-coeffs-zoom-whirl-zooming}}
\end{figure}

Using the vector potential code described above I can reproduce the results
obtained from the Faraday tensor method discussed in the main paper. The
differences are small, typically of the order of $10^{-3}\%$ of the field
values as shown in Fig.~\ref{fig:vector-faraday-comparison}. 
I expect the Faraday tensor code to yield more
accurate results since the costly numerical differentiation that is necessary
in the vector potential calculation is absent. Nevertheless I can
reproduce \eg\ the correct decay behaviour of the multipole coefficients for
a zoom-whirl orbit as shown in
Fig.~\ref{fig:vector-multipole-coeffs-zoom-whirl-zooming}.

\section{Jump conditions}
\subsection{Faraday tensor calculation\label{sec:jump-conditions}}
Since the source term in Eqs.~\eqref{eqn:red-wave-psi} --
\eqref{eqn:red-wave-xi} contains a term proportional to $\Dirac'(r^*-r^*_0)$, the
field is discontinuous across the world line of the particle.
I use 
\begin{multline}
    \jump{\partial_t^n\partial_{r^*}^m\psi} =
    \lim_{\varepsilon \approaches 0^+} 
    [\partial_t^n\partial_{r^*}^m\psi(t_0,r_0^*+\varepsilon) 
    \\
    - \partial_t^n\partial_{r^*}^m\psi(t_0, r_0^*-\varepsilon)]
\end{multline}
to denote the jump in $\partial^n_t \partial^m_{r^*} \psi$ across the world
line. I only calculate jump conditions in the $r^*$ direction
up to $\jump{\partial_{r^*} \psi}$, which I find by substituting the ansatz
\begin{align}
    \psi &= 
    \psi_{<}(t, r^*) \Heaviside(r^*_0-r^*) 
    \nonumber\\&\quad
    + \psi_{>}(t, r^*) \Heaviside(r^*-r^*_0)
\end{align}
into Eqs.~\eqref{eqn:red-wave-psi} --
\eqref{eqn:red-wave-xi} and its $t$ and $r^*$ derivatives.
Demanding in each step that the singularity structure on the left hand side
matches that of the sources (and their derivatives) on the right hand side
yields the jump conditions
\begin{align}
    \jump{\psi} &= \frac {F_\psi}{f_0 [(\partial_t r^*_0)^2-1]}
\text{,}\\
\intertext{and}
    \jump{\partial_{r^*} \psi} &=- \frac {G_\psi}{(\partial_t r^*_0)^2-1}
    \nonumber\\&\quad
    - \frac {\partial_t^2 r^*_0 \left[ 3\,(\partial_t r^*_0)^2+1 \right] F_\psi}{f_0\,[(\partial_t r^*_0)^2-1]^3}
    \nonumber\\&\quad
    +2 \frac {\partial_t r^*_0\,\partial_t \left( F_\psi/f_0 \right)}{[(\partial_t r^*_0)^2-1]^2}
%\text{,}\\
%\jump{\partial_{r^*}^2 \psi} &= \frac {\partial_t^2 r^*_0 [3\,(\partial_t r^*_0)^{2}+1] G_\psi}{[(\partial_t r^*_0)^2-1]^3}
%    -2 \frac {\partial_t r^*_0\,\partial_t G_\psi}{ [(\partial_t r^*_0)^2-1]^2}
%    \nonumber\\&\mbox{}
%    + \left\{ 3 \frac { \left[ 10\,(\partial_t r^*_0)^{2}+1+5\,(\partial_t r^*_0)^4 \right] (\partial_t^2 r^*_0)^2}{[(\partial_t r^*_0)^2-1]^5}
%      - \frac {V}{ [(\partial_t r^*_0)^2-1]^2}
%      -4 \frac {\partial_t r^*_0\, \left[ (\partial_t r^*_0)^2+1 \right] \partial_t^3 r^*_0}{[(\partial_t r^*_0)^2-1]^4}
%    \right\} F_\psi/f_0
%    \nonumber\\&\mbox{}
%    -12  \frac {\partial_t r^*_0\, \left[(\partial_t r^*_0)^2+1\right] \partial_t^2 r^*_0\, \partial_t \left( F_\psi/f_0 \right)}{[(\partial_t r^*_0)^2-1]^4}
%    + \frac { \left[ 3(\partial_t r^*_0)^2+1 \right] \partial_t^2 \left( F_\psi/f_0 \right)}{ [(\partial_t r^*_0)^2-1]^3}
%    \nonumber\\&\mbox{}
%    -\left\{ \frac {V_\xi F_\psi}{ f_0 [(\partial_t r^*_0)^2-1]^2} \text{ for $\jump{\partial_{r^*}^2\xi}$}\right\}
\text{,}
\end{align}
where $\psi$ stands for either one of $\psi$, $\chi$, or $\xi$. 
%$f_0 = 1-\frac{2M}{r_0}$, $V = \ell(\ell+1)\frac{r_0-2M}{r_0^3}$.
%, $V_\xi = 2\frac{(r_0-2M)(r_0-3M)}{r_0^5}$. The last term involving $V_\xi F_\psi$ only
%appears for $\jump{\partial_{r^*}^2\xi}$.

\subsection{Vector potential calculation\label{sec:vector-jump-conditions}}
Since the source term in Eq.~\eqref{eqn:covariant-wave-eqn} is singular, the
field is only continuous across the world line of the particle, but not
smooth. 
I use 
\begin{multline}
    \jump{\partial_t^n\partial_{r^*}^m\psi} =
    \lim_{\varepsilon \approaches 0^+} 
    [\partial_t^n\partial_{r^*}^m\psi(t_0,r_0^*+\varepsilon) 
    \\
    - \partial_t^n\partial_{r^*}^m\psi(t_0, r_0^*-\varepsilon)]
\end{multline}
to denote the jump in $\partial^n_t \partial^m_{r^*} \psi$ across the world
line. For my purposes I only need the jump conditions in the $r^*$ direction
up to $\jump{\partial^2_{r^*} \psi}$, which I find by substituting the ansatz
\begin{align}
    A^{\ell m}_a(t, r^*) &= 
    A^{\ell m}_{a,<}(t, r^*) \Heaviside(r^*_0-r^*) 
    \nonumber\\&\quad
    + A^{\ell m}_{a,>}(t, r^*) \Heaviside(r^*-r^*_0)
    \text{,}
    \\
    v^{\ell m}(t, r^*) &= 
    v^{\ell m}_{<}(t, r^*) \Heaviside(r^*_0-r^*) 
    \nonumber\\&\quad
    + v^{\ell m}_{>}(t, r^*) \Heaviside(r^*-r^*_0)
    \text{,}
    \\
    \tilde v^{\ell m}(t, r^*) &= 
    \tilde v^{\ell m}_{<}(t, r^*) \Heaviside(r^*_0-r^*) 
    \nonumber\\&\quad
    + \tilde v^{\ell m}_{>}(t, r^*) \Heaviside(r^*-r^*_0)
\end{align}
into Eqs.~\eqref{eqn:reduced-wave-eqn-begin} --
\eqref{eqn:reduced-wave-eqn-end} and its $t$ and $r^*$ derivatives.
Demanding in each step that the singularity structure on the left hand side
matches that of the sources (and their derivatives) on the right hand side
yields the jump conditions
\begin{gather}
    \jump{A^{\ell m}_a} = \jump{w^{\ell m}} = 0
    \text{,}
    \\
    \jump{\partial_{r^*} A^{\ell m}_a} = \frac{E^2}{E^2-\dot r_0^2} S_a
    \text{,}
    \\
    \jump{\partial_{r^*} w^{\ell m}} = \frac{E^2}{E^2-\dot r_0^2} S_{\text{even/odd}}
    \text{,}
    \\
    \begin{split}
    \jump{\partial^2_{r^*} A^{\ell m}_a} = 
    \left( \frac{2 M E^4}{r_0^2 (E^2-\dot r_0^2)^2}- f_0 
      \frac{(3\dot r_0^2 +E^2) E^2 \ddot r_0}{(E^2-\dot r_0^2)^3} \right) S_a 
      \\\mbox{}
      + \frac{2M E^3 \dot r_0}{r_0^2 (E^2-\dot r_0^2)^2} S_b -
      f_0 \frac{2 E^2 \dot r_0}{(E^2-\dot r_0^2)^2}\dot S_a \\ 
    \text{,}
    \end{split}
    \\
    \begin{split}
    \jump{\partial^2_{r^*} w^{\ell m}} = -f_0
      \frac{(3\dot r_0^2+E^2) E^2 \ddot r_0}{(E^2-\dot r_0^2)^3} S_{\text{even/odd}} 
      \\\mbox{}
      - f_0 \frac{2E^2 \dot r_0}{(E^2-\dot r_0^2)^2} \dot S_{\text{even/odd}} 
      \text{,}
    \end{split}
\end{gather}
where $a,b \in \{t, r^*\}, a \ne b$, $w \in \{v,
\tilde v\}$.

%%fakesection
\bibliography{paper}

\end{document}